\documentstyle[aps,prd,eqsecnum,amssymb,amsfonts,preprint,tighten,%
floats,epsfig]{revtex}
\newcommand{\mc}[1]{{\cal #1}}
\newcommand{\abs}[1]{\left|#1\right|}
\newcommand{\dual}[1]{{}^*\!\left(#1\right)}
\newcommand{\eqref}[1]{(\ref{#1})}
\newcommand{\R}{\mathbb{R}}
\newcommand{\coord}{coordinate}
\newcommand{\coeff}{coefficient}
\newcommand{\mink}{Min\-kow\-ski}

\begin{document}
\draft
\title{Quasi-stationary binary inspiral. I.
  Einstein equations for the two Killing vector
  spacetime\thanks{Published as \textit{Phys Rev D} \textbf{60},
    084009 (1999).  Copyright 1999 by the American Physical Society}}
\preprint{gr-qc/9812041}
\author{John T.~Whelan\thanks{Electronic address:
    whelan@itp.unibe.ch}}
\address{Institut f\"{u}r theoretische Physik, Universit\"{a}t Bern,
  Sidlerstrasse 5, CH-3012 Bern, Switzerland}
\author{Joseph D.~Romano\thanks{Electronic address: 
    jromano@utb1.utb.edu}}
\address{Department of Physical Sciences, 
  University of Texas at Brownsville, Brownsville, Texas 78521}
\date{April 13, 1999}

\maketitle

\begin{abstract}
  The inspiral of a binary system of compact objects due to
  gravitational radiation is investigated using the toy model of two
  infinitely long lines of mass moving in a fixed circular orbit.  The
  two Killing fields in the toy model are used, according to a
  formalism introduced by Geroch, to describe the geometry entirely in
  terms of a set of tensor fields on the two-manifold of Killing
  vector orbits.  Geroch's derivation of the Einstein equations in
  this formalism is streamlined and generalized.  The explicit
  Einstein equations for the toy model spacetime are derived in terms
  of the degrees of freedom which remain after a particular choice of
  gauge.
\end{abstract}

\narrowtext

\section{Introduction}

A pair of compact objects (black holes or neutron stars) in binary
orbit about one another is stable in Newtonian gravity.  In general
relativity, however, the system will emit gravitational radiation,
causing the bodies to spiral in towards one another.  The early stages
of this process, where the gravitational interaction is weak
everywhere, can be treated with the post-Newtonian approximation, while the
final merger of the objects can be modeled using supercomputers.  In
order to determine the waveform in an intermediate phase, where the
rate of energy loss due to gravitational radiation is low, but other
strong-field effects may be important, and also to provide accurate
initial data for supercomputer calculations, it is useful to employ an
approximation scheme based on the fact that the orbits are decaying
only slowly.  Over some range of time, the physical spacetime should
be approximated by a spacetime in which the orbits do not
decay \cite{det}.\footnote{To maintain this equilibrium, the energy lost in
  gravitational waves is presumably balanced by gravitational
  radiation coming in from infinity \cite{standing}.}   For
elliptical orbits this spacetime will be periodic, with the period
equal to the orbital period of the objects.  If the orbits are
circular, this discrete symmetry becomes a continuous symmetry and the
spacetime is stationary.  Finding this spacetime is thus an
essentially three-dimensional problem, rather than a four-dimensional
one.

Solving this three-dimensional problem is still numerically intensive,
however, so in order to examine the consequences of this approximation
scheme in a computationally simpler environment, we consider initially
a toy model which exhibits an additional translational symmetry
perpendicular to the orbital plane.  The desired spacetime then has
two Killing vectors, and we have only a two-dimensional problem.

This, the first in a series of papers on this project, is concerned
with formulating the Einstein equations in the presence of these two
Killing symmetries.  Paper II \cite{standing} examines, in the simpler
model of a non-linear scalar field in $2+1$ dimensions, the details of
the radiation-balanced boundary conditions which must be imposed ``at
infinity'' to specify a stationary, energy-conserving solution to a
radiative system.  Subsequent papers will apply this method of
radiation balance to the gravitational fields of co-orbiting lines of
mass and ultimately to localized sources, thus removing one at a time
the simplifications of scalar field theory and translational
invariance.

In this paper, we describe a toy model for binary inspiral that has
such a two-dimensional symmetry group.  The model problem consists of
two infinitely long lines of mass (e.g., cosmic strings) orbiting one
another at fixed angular velocity.  Using a formalism introduced by
Geroch \cite{GEROCH2}, we describe the geometry of the toy model
spacetime in terms of a set of tensor fields on the two-dimensional
manifold of Killing vector orbits.  We also derive explicit
expressions for the Einstein equations for the spacetime in terms of
the degrees of freedom which remain after a particular choice of
gauge.  Future work will solve these equations numerically for a
reasonable set of boundary conditions.

The plan of this paper is as follows:

In Sec.~\ref{sec:problem}, we begin our analysis by describing
Minkowski spacetime in a ``co-rotating'' {\coord} system corresponding
to two co-orbiting cosmic strings.  We show that, even for this simple
spacetime, the two Killing vector fields (KVFs) defined by the strings
do \emph{not} select a preferred {\coord} system consisting of two
Killing {\coord}s and two {\coord}s on an orthogonal subspace.

In Sec.~\ref{sec:2KVF}, we describe a formalism which can be used to
simplify the discussion of spacetimes with two KVFs, even in the
absence of a system of mutually orthogonal Killing and non-Killing
{\coord}s.  The formalism was initially developed by Geroch
\cite{GEROCH2}, and we generalize his derivation of the vacuum
Einstein equations by deriving expressions for different projections
of the Einstein tensor.  We also show how the various components of
the Einstein tensor are related by the contracted Bianchi identities,
and we show explicitly how to recover the four-geometry from Geroch's
more specialized objects.  The discussion in Sec.~\ref{sec:2KVF}
applies to \emph{any} spacetime with two commuting Killing vector
fields; it is \emph{not} specialized to the co-rotating cosmic string
spacetime considered in the rest of the paper.

In Sec.~\ref{sec:fix}, we return our focus to the co-rotating cosmic
string spacetime by discussing the gauge choices available within
Geroch's formalism.  We describe some desirable gauge-fixings, and we
enumerate the independent degrees of freedom which remain.

In Sec.~\ref{sec:PDEs}, we derive explicit expressions for the
components of the Einstein tensor in terms of the independent
functions needed to describe the geometry.  These equations, when
supplemented by a description of the stress-energy of the cosmic
strings and a set of boundary conditions \cite{standing,cyl}, set the
stage for a numerical solution of the Einstein equations, which will
be performed in the future.

Finally, in Sec.~\ref{sec:concl}, we summarize the results of our paper
and discuss how they will be used as the starting point for future work.

The first two appendices contain proofs of formulas used in
Sec.~\ref{sec:2KVF}: Appendix~\ref{app:grad_K} contains a proof of the
expression (\ref{e:grad_K}) for the covariant derivative of a Killing
vector field, and Appendix~\ref{app:ricci} contains a detailed
derivation of the projected components of the Ricci tensor,
which we state 
in Sec.~\ref{sec:2KVF}.  The third and fourth appendices consider
ancillary subjects: Appendix~\ref{app:coord} describes how to obtain
from the quantities defined by Geroch the components of the
four-metric in a {\coord} basis rather than the non-{\coord} one used
in the text, and recovers the normal form given by Petrov
\cite{petrov} as a consequence of a particular {\coord} choice.
Finally, Appendix~\ref{app:geroch} demonstrates the correspondence
between the formulas contained in Sec.~\ref{sec:2KVF} and those given
by Geroch in Appendix~A of \cite{GEROCH2}, the latter being a special
case of the former.

Note: Throughout this paper, we will follow the sign conventions of
\cite{MTW}.  Abstract indices are denoted by lower case Latin letters
$a,b,\ldots$\ from the beginning of the alphabet, while spacetime
{\coord} indices are denoted by lower case Greek letters
$\mu,\nu,\ldots $\ .  The Killing vectors are labeled by upper case
Latin letters $A,B,\ldots$\ , and the two-dimensional {\coord} indices
on the space of orbits of the Killing vectors by lower case Latin
letters $i,j,\ldots$\ from the middle of the alphabet.

\section{Minkowski spacetime in co-rotating coordinates}
\label{sec:problem}

As mentioned in the previous section, we wish to describe a spacetime
which has two infinitely long cosmic strings%
\footnote{These are cosmic strings with an actual curvature
  singularity, and not only a deficit angle.  Two conical
  singularities surrounded by flat spacetime could scatter
  gravitationally, but could not orbit one another.}  orbiting one
another at a fixed angular velocity $\Omega$.  In a numerical
determination of the spacetime geometry, one seeks to fix the {\coord}
(i.e., gauge) information completely, and thus calculate the minimum
number of quantities necessary to define the geometry.  It is
desirable, of course, to choose a gauge which takes advantage of the
symmetries of the problem.  In this case, those symmetries are
described by two KVFs.  One of these, $K_1^a$, corresponds to the
translational invariance along the strings, while the other, $K_0^a$,
tells us that the spacetime is unchanged if we move forward in time
while rotating about the axis by a proportional amount.  The desired
{\coord} system would seem to consist of Killing {\coord}s $x^0$ and
$x^1$, supplemented by {\coord}s $x^2$ and $x^3$ on an orthogonal
subspace.  (Indeed, this is what is done \cite{wald} in the case of
stationary, axisymmetric spacetimes, which also admit two commuting
Killing vectors.)  However, in the case at hand, the ``co-rotational''
Killing vector $K_0^a$ is not surface-forming, and \emph{there is no
  subspace orthogonal to the Killing vectors}.  This is illustrated by
describing {\mink} spacetime in {\coord}s tailored to the symmetries
exhibited by co-rotating cosmic strings.

The {\mink} metric $g_{ab}$, written in standard cylindrical polar
{\coord}s $(t,z,\rho,\phi)$, gives rise to the line element
\begin{equation}
  ds^2=g_{\mu\nu}dx^\mu dx^\nu=-dt^2+dz^2+d\rho^2+\rho^2\ d\phi^2
  .
\end{equation}
We can work in a reference frame which rotates with a fixed angular
velocity $\Omega$ by defining a ``co-rotating'' angle
\begin{equation}
  \varphi:=\phi - \Omega t
\end{equation}
and transforming to {\coord}s $(t,z,\rho,\varphi)$.  In these
{\coord}s the line element takes the form
\begin{equation}
  ds^2=-dt^2+dz^2+d\rho^2+\rho^2(d\varphi+\Omega dt)^2
  ,
\end{equation}
which can be expanded to yield
\begin{equation}
  \label{rotline}
ds^2=
-(1-\Omega^2\rho^2)dt^2+dz^2+d\rho^2+\rho^2\ d\varphi^2
  +2\Omega\rho^2 d\varphi\ dt
  .
\end{equation}
%

If we limit consideration of the symmetries of the spacetime to those 
described by the two commuting KVFs
\begin{mathletters}
  \begin{eqnarray}
    K_0^a&:=&
    \left(
      \frac{\partial}{\partial t}
    \right)_\varphi
    \equiv\left(
      \frac{\partial}{\partial t}
    \right)_\phi
    +\Omega\frac{\partial}{\partial\phi}
    ;
    \\
    K_1^a&:=&\frac{\partial}{\partial z}
    ,
  \end{eqnarray}
\end{mathletters}
we see that $t$ and $z$ are the corresponding Killing {\coord}s for
the metric written in the form \eqref{rotline}.  The presence of a
$d\varphi\ dt$ term in Eq.~\eqref{rotline} means that the {\coord} pairs
$(t,z)$ and $(\rho,\varphi)$ have not split the spacetime into
orthogonal subspaces.%
\footnote{A division based on the {\coord} pairs $(t,\varphi)$ and
  $(\rho,z)$ \emph{does} split the spacetime into orthogonal
  subspaces.  However, although $\varphi$ is a Killing {\coord} in
  {\mink} space, it will not be in the cosmic string spacetime with
  which we are concerned, and so such a {\coord} splitting is not of
  interest to us.}  In fact it is \emph{impossible} to base such a
split on these two Killing vectors, for while $K_1^a$ is clearly
surface forming, $K_0^a$ is not, as calculation of
\begin{equation}
  \label{hyper}
  \epsilon^{abcd}K_{0b}\nabla_c K_{0d}
\end{equation}
clearly shows.%
\footnote{In the language of differential forms,
  $K_0=-dt+\Omega\rho^2 d\phi$ and $K_0\wedge d K_0=-2\Omega\rho\ 
  dt\wedge d\rho\wedge d\phi$.  Equivalently, 
  $\dual{K_0\wedge d K_0}=2\Omega dz$, where $*$ denotes the duality
  operator (see, e.g., p.~88 of \cite{wald}).}  
The other Killing vector might save us, if Eq.~\eqref{hyper} had
a vanishing projection along $K_1^a$, but since%
\footnote{Again, working
  with differential forms, $K_1=dz$ and $ K_0\wedge
  K_1\wedge\ d K_0= -2\Omega\rho\ dt\wedge dz\wedge d\rho\wedge d\phi$.
  Equivalently, $\dual{K_0\wedge K_1\wedge\ d K_0}=c_0=2\Omega$.}
\begin{equation}
  c_0:=\epsilon^{abcd}K_{0a}K_{1b}\nabla_c K_{0d}=2\Omega\ne 0
  ,
\end{equation}
the group of symmetries is not \emph{orthogonally transitive} and the
two-dimensional subspaces of the tangent space at each point orthogonal
to $K_0^a$ and $K_1^a$ are not integrable.

If the symmetry group were orthogonally transitive, we could define a
{\coord} system made of two Killing {\coord}s $\{x^A|A=0,1\}$ and two
{\coord}s $\{x^i|i=2,3\}$ on an orthogonal subspace.  In that case,
the metric would be block diagonal and defined by two $2\times 2$
symmetric matrices: (i) the matrix of inner products
\begin{equation}
  \lambda_{AB}:=g_{ab}K_A^a K_B^b
\end{equation}
and (ii) the $ij$-components of the projection
tensor\footnote{$\{\lambda^{AB}\}$ is the matrix inverse of
  $\{\lambda_{AB}\}$.}
\begin{equation}
  \gamma_{ab}:=g_{ab}-\lambda^{AB}K_{Aa}K_{Bb}
  .
  \label{e:equation1}
\end{equation}
Examination of the metric \eqref{rotline} shows that the matrix of
components $\{g_{\mu\nu}\}$ [with respect to the co-rotating {\coord}s
$\{x^\mu\}:=\{x^A;x^i\}=(t,z;\rho,\varphi)$] is not block-diagonal, so
that
\begin{equation}
  ds^2=g_{\mu\nu}dx^\mu dx^\nu\ne\lambda_{AB}dx^A dx^B
  +\gamma_{ij}dx^i dx^j
  .
\end{equation}
However, the quantities $\{\lambda_{AB}\}$ and $\gamma_{ab}$ are still
useful in the construction in Sec.~\ref{sec:2KVF}, so we will examine
their form for co-rotating flat spacetime to keep them in mind as an
example.

The matrix of inner products is
\begin{equation}
\label{lambdaAB}
  \{\lambda_{AB}\}=
  \left(
    \begin{array}{cc}
      -(1-\Omega^2\rho^2) & 0     \\
      0                   & 1
    \end{array}
  \right)
  .
\end{equation}
The determinant
\begin{equation}
  \label{corotlambda}
  \lambda:=\det\{\lambda_{AB}\}=-(1-\Omega^2\rho^2)
\end{equation}
is less than zero for $\rho<1/\Omega$,
greater than zero for $\rho>1/\Omega$, and
equal to zero for $\rho=1/\Omega$.  The surface on which
$K_0^a=(\partial/\partial t)_\varphi$ is null, defined by
$\rho=1/\Omega$, is known as the ``light cylinder.''%
\footnote{This is because an object which sat at
  $\rho=1/\Omega$, with constant $z$ and $\varphi$, would be moving at
  the speed of light.}
For $\rho<1/\Omega$, $K_0^a$ is timelike, while
for $\rho>1/\Omega$, $K_0^a$ is spacelike. 

In terms of $\{x^\mu\}=(t,z,\rho,\varphi)$, 
the projection tensor $\gamma_{ab}$ has components
\begin{equation}
\{\gamma_{\mu\nu}\}=
\left(
\begin{array}{cccc}
0       & 0     & 0     & 0     \\
0       & 0     & 0     & 0     \\
0       & 0     & 1     & 0     \\
0       & 0     & 0     & \rho^2(1-\Omega^2\rho^2)^{-1}
\end{array}
\right)
.
\end{equation}
As we shall describe in Sec.~\ref{sec:2KVF}, $\gamma_{ab}$ can be
thought of as a metric on the space $\mc{S}$ of Killing vector orbits,
whose line element [in terms of the {\coord}s
$\{x^i\}:=(\rho,\varphi)$] is
\begin{equation}
  \label{corotmet}
  d\Sigma^2:=\gamma_{ij}dx^i dx^j=d\rho^2 
  + \rho^2(1-\Omega^2\rho^2)^{-1}\   
  d\varphi^2
  .
\end{equation}
Note that this metric has signature $(++)$ for $\rho<1/\Omega$,
$(+-)$ for $\rho>1/\Omega$, and is degenerate for $\rho=1/\Omega$.
The light cylinder $\rho=1/\Omega$ can thus be thought of as a 
``signature change surface'' in $\mc{S}$.
The determinant 
\begin{equation}
\gamma:=\det\{\gamma_{ij}\}=\rho^2(1-\Omega^2\rho^2)^{-1}
\end{equation}
diverges when $\rho=1/\Omega$, which is exactly when the matrix
$\{\lambda_{AB}\}$ becomes non-invertible.

\section{Spacetimes with two commuting Killing vector fields}
\label{sec:2KVF}

In this section, we describe a general formalism (originally developed
by Geroch \cite{GEROCH2}) that can be used to simplify the discussion
of spacetimes admitting two commuting KVFs, even in the absence of of
a system of mutually orthogonal Killing and non-Killing coordinates.
We present a new derivation of the projected form of the Einstein
equations (and the Bianchi identities which relate various components
of the Einstein tensor), and we show how to reconstruct the original
four-geometry, given only the values of certain tensor fields on the
two-dimensional space of Killing vector orbits.  The analysis that we
give in this section is completely general.  In particular, we do not
restrict attention to the case of the two co-rotating cosmic string
spacetime, which we consider in the rest of the paper.

\subsection{Preliminaries}
\label{ssec:prelim}

Let $(\mc{M},g_{ab})$ be a four-dimensional manifold $\mc{M}$ with
Lorentzian metric $g_{ab}$, which admits two commuting Killing vector
fields $K_A^a$ $(A=0,1)$.  Killing's equation $\mc{L}_{K_A}g_{ab}=0$
is equivalent to
\begin{equation}
\nabla_a K_{Ab}=-\nabla_b K_{Aa}\ ,
\label{e:KVF}
\end{equation}
while commutivity of the vector fields $[K_A,K_B]^a=0$ is equivalent
to
\begin{equation}
K_A^b\nabla_b K_B^a=K_B^b\nabla_b K_A^a\ .
\label{e:commute}
\end{equation}
In addition,
\begin{equation}
  \label{e:Riem_K}
  R^a{}_{bcd}K_{Aa}
  =\nabla_b\nabla_c K_{Ad}
  \ ,
\end{equation}
which is valid for any Killing vector.  Since we will not assume that
the KVFs are orthogonally transitive (i.e., that the two-dimensional
subspaces orthogonal to $K_0^a$ and $K_1^a$ are integrable), one or
both of the quantities
\begin{equation}
c_A:=\epsilon_{abcd}K_0^a K_1^b\nabla^c K_A^d
\label{e:c_A}
\end{equation}
can be non-zero somewhere in $\mc M$.

Given $g_{ab}$ and $K_A^a$, we can construct the symmetric matrix
of inner products
\begin{equation}
\lambda_{AB}:=g_{ab}K_A^a K_B^b\ .
\label{e:lambda_AB}
\end{equation}
If the determinant
\begin{equation}
\lambda:={\rm det}\left\{\lambda_{AB}\right\}
\label{e:det}
\end{equation}
is non-zero,
then we can further define a projection tensor
\begin{equation}
\gamma_{ab}:=g_{ab}-\lambda^{AB} K_{Aa} K_{Bb}\ ,
\label{e:gamma_ab}
\end{equation}
where $\{\lambda^{AB}\}$ denotes the inverse matrix to
$\{\lambda_{AB}\}$.  $\gamma_{ab}$ is orthogonal to the KVFs, and it
can be interpreted as a metric on the two-dimensional space $\mc{S}$
of Killing vector orbits.  In fact, as shown by Geroch \cite{GEROCH2},
{\em any} tensor field $T^{a_1\cdots a_n}{}_{b_1\cdots b_m}$ on
$\mc{M}$ that: (i) is orthogonal to the KVFs
\begin{eqnarray}
K_{Aa_1}T^{a_1\cdots a_n}{}_{b_1\cdots b_m}=0\ ,\ 
&\ldots&\ ,\ 
K_{Aa_n}T^{a_1\cdots a_n}{}_{b_1\cdots b_m}=0\ ,
\nonumber\\
K_{A}^{b_1}T^{a_1\cdots a_n}{}_{b_1\cdots b_m}=0\ ,\ 
&\ldots&\ ,\ 
K_{A}^{b_m}T^{a_1\cdots a_n}{}_{b_1\cdots b_m}=0\ ,
\nonumber\\
\label{e:orthog}
\end{eqnarray}
and (ii) has vanishing Lie derivatives
\begin{equation}
\mc{L}_{K_A}T^{a_1\cdots a_n}{}_{b_1\cdots b_m}=0
\label{e:Lie_drag}
\end{equation}
can be thought of as a tensor field on $\mc{S}$.
In particular, since 
\begin{equation}
\mc{L}_{K_C}\lambda_{AB}=0\quad{\rm and}\quad
\mc{L}_{K_C} c_A=0\ ,
\label{e:Lie_lambda}
\end{equation}
$\lambda_{AB}$ and $c_A$ are scalar fields on $\mc{S}$.

The metric-compatible covariant derivative operator $D_a$ 
on $\mc{S}$ is given by
\begin{equation}
\label{e:D_a}
D_a T^{b_1\cdots b_n}{}_{c_1\cdots c_m}:=
\gamma_a^d\gamma^{b_1}_{e_1}\cdots\gamma^{b_n}_{e_n}
\gamma_{c_1}^{f_1}\cdots\gamma_{c_m}^{f_m}
\nabla_d T^{e_1\cdots e_n}{}_{f_1\cdots f_m}\ ,
\end{equation}
where $T^{a_1\cdots a_n}{}_{b_1\cdots b_m}$ is any tensor field on
$\mc{M}$ satisfying Eqs.~(\ref{e:orthog}) and (\ref{e:Lie_drag}), and the
two-dimensional Levi-Civita tensor $\epsilon_{ab}$ can be written as
\begin{equation}
\epsilon_{ab}=\abs{\lambda}^{-1/2}\epsilon_{abcd}K_0^c K_1^d\ .
\label{e:eps_ab}
\end{equation}
If we define $\epsilon$ in terms of $\lambda$ and its absolute value
via
\begin{equation}
  \label{epsilon_def}
\lambda=\epsilon\abs{\lambda}
\end{equation}
(so that $\epsilon=1$ corresponds to two spacelike KVFs, and 
$\epsilon=-1$ to one spacelike and one timelike KVF),
then
\begin{equation}
  \epsilon^{ab}\epsilon_{ac}=-\epsilon\delta_c^b
  .
\end{equation}
Moreover, if we define a Levi-Civita symbol $\epsilon^{AB}$ so that
$\epsilon^{01}=\epsilon^{-1}\abs{\lambda}^{-1/2}$, then 
Eqs.~\eqref{e:c_A} and \eqref{e:eps_ab} can be rewritten as
\begin{equation}
c_A=\frac{1}{2}\epsilon\abs{\lambda}^{1/2}
\epsilon_{abcd}\epsilon^{CD}K_C^a K_D^b\nabla^c K_A^d
\label{e:c_A_new}
\end{equation}
and
\begin{equation}
  \epsilon_{ab}
  =\frac{1}{2}\epsilon\epsilon_{abcd}\epsilon^{AB}K_A^cK_B^d
  \ ,
\label{e:eps_ab_new}
\end{equation}
respectively, which do not explicitly involve the indices 0 and 1.
The presence of $\abs{\lambda}^{1/2}$ in Eq.~\eqref{e:c_A_new} implies
that $c_A$ transforms as a covariant vector \emph{density} of weight
$+1$, rather than as a covariant vector, with respect to the index $A$.
Similarly, the absence of any $\lambda$ factors in
Eq.~\eqref{e:eps_ab_new} implies that $\epsilon_{ab}$ carries no density
weight.

\subsection{Projected Ricci tensor}
\label{ssec:proj_ricci}

Given the definitions of the previous subsection, we are now ready to
calculate the projected components of the four-dimensional Ricci
tensor $R_{bd}:=R^c{}_{bcd}$.  This is the first (and most involved)
step leading to the projected form of the Einstein equations
$G_{ab}=8\pi T_{ab}$.  Since
\begin{equation}
G_{ab}:=R_{ab}-\frac{1}{2}R g_{ab}\ ,
\label{e:G_ab}
\end{equation}
it follows that 
\begin{equation}
R_{ab}=8\pi\left(T_{ab}-\frac{1}{2}T g_{ab}\right)
\ ,
\label{e:EE2}
\end{equation}
where $T:=T_{ab} g^{ab}$ denotes the trace of the stress-energy tensor.
Thus, knowing the projections 
\begin{mathletters}
\begin{eqnarray}
R_{AB}&:=&K_A^c K_B^d R_{cd}\\
\widehat{R}_{Ab}&:=&K_A^c \gamma_b^d R_{cd}\\
\widehat{R}_{ab}&:=&\gamma_a^c \gamma_b^d R_{cd}
\end{eqnarray}
\end{mathletters}
of the Ricci tensor is equivalent to knowing the projections of the 
left-hand side of the Einstein equations.

In \cite{GEROCH2}, Geroch derived the projected form of the Einstein 
equations for vacuum spacetimes admitting one timelike and one 
spacelike Killing vector field.
In this paper, we extend Geroch's derivation in the following ways:

(i) we consider non-vacuum spacetimes by allowing a non-zero
stress-energy tensor $T_{ab}$;

(ii) we allow the KVFs to have either ``signature''---i.e., they can
both be spacelike ($\epsilon=1$), or one can be spacelike and the
other timelike ($\epsilon=-1$);

(iii) we take advantage of the index notation to treat both KVFs 
simultaneously.

In addition, our derivation is somewhat simpler than Geroch's in the
sense that, instead of introducing the symmetric matrix of twist
vectors $\omega_{AB}^a$ [(A8) of \cite{GEROCH2}] and their projections
$\nu_{AB}^a$ [(A10) of \cite{GEROCH2}]\footnote{These quantities are
  considered in Appendix~\ref{app:geroch} for the sake of identifying
  out results with Geroch's.}, we make repeated use of the expression
\begin{equation}
\nabla_a K_{Ab}=-\frac{1}{2}\epsilon^{-1}\abs{\lambda}^{-1/2}
\epsilon_{ab}\ c_A
-\lambda^{BC}K_{B[a}D_{b]}\lambda_{CA}
\label{e:grad_K}
\end{equation}
for the covariant derivative of the KVFs.
[The use of Eq.~(\ref{e:grad_K}) \emph{greatly} simplifies calculations
involving one or two derivatives of the KVFs.]
A proof of Eq.~(\ref{e:grad_K}) can be found in Appendix~\ref{app:grad_K}.
  
The projected components of the four-dimensional Ricci tensor are
worked out in detail in Appendix~\ref{app:ricci}.  The final results,
which we simply state here, are
\begin{mathletters}
  \label{e:R}
  \begin{eqnarray}
    \label{e:RKK}
    R_{AB}&=&
    -\frac{1}{2}D^aD_a\lambda_{AB}
    +{1\over 4}(\lambda^{-1}D^a\lambda)D_a\lambda_{AB}
    \nonumber
    \\
    &&-{1\over 4}\lambda_{AB}\lambda^{-1}(D^a\lambda\lambda^{CD})D_a
    \lambda_{CD}-\frac{1}{2}\lambda^{-1}c_A c_B
    ;
    \\
    \label{e:RKg}
    \widehat{R}_{Ab}&=&
    -\frac{1}{2}\epsilon^{-1}\abs{\lambda}^{-1/2}\epsilon_{bc}D^cc_A
    ;
    \\
    \nonumber
    \widehat{R}_{ab}&=&
    -\frac{1}{2}\lambda^{-1}D_aD_b\lambda
    +\frac{1}{4}\lambda^{-1}(D_a\lambda\lambda^{AB})(D_b\lambda_{AB})
    \\
    &&+\frac{1}{4}\lambda^{-2}(D_a\lambda)(D_b\lambda)
    +\frac{1}{2}\gamma_{ab}
    (\mc{R}+\lambda^{-1}\lambda^{AB}c_A c_B)
    .
    \label{e:Rgg}
  \end{eqnarray}
\end{mathletters}

\subsection{Trace and trace-free parts of the projected Ricci tensor}
\label{ssec:trace}

For reasons which shall become clearer in Sec.~\ref{sec:PDEs}, it is
convenient to split the projected components $R_{AB}$ and
$\widehat{R}_{ab}$ of the Ricci tensor into their trace and trace-free
parts with respect to $\{\lambda^{AB}\}$, $\gamma^{cd}$, and the
projection operators
\begin{mathletters}
\begin{eqnarray}
P_{AB}^{CD}&:=&\delta_{(A}^C\delta_{B)}^D
-\frac{1}{2}\lambda_{AB}\lambda^{CD}\ ,\\
P_{ab}^{cd}&:=&\gamma_{(a}^c\gamma_{b)}^d
-\frac{1}{2}\gamma_{ab}\gamma^{cd}\ .
\end{eqnarray}
\end{mathletters}
The projection operators satisfy
\begin{mathletters}
  \label{e:proj_ids}
  \begin{eqnarray}
    &&P_{AB}^{CD}P_{CD}^{EF}=P_{AB}^{EF}\ ,
    \quad\quad
    P_{ab}^{cd}P_{cd}^{ef}=P_{ab}^{ef}
    \ ,\\
    &&\lambda^{AB}P_{AB}^{CD}=
    P_{AB}^{CD}\lambda_{CD}=
    \gamma^{ab}P_{ab}^{cd}=
    P_{ab}^{cd}\gamma_{cd}=0
    \ ,\\
    &&P_{ab}^{cd}K_A^a=
    P_{ab}^{cd}K_A^b=
    P_{ab}^{cd}K_{Ac}=
    P_{ab}^{cd}K_{Ad}=0
    .
  \end{eqnarray}
\end{mathletters}
Using these results together with Eqs.~\eqref{e:R} and \eqref{lDDl}, it 
immediately follows that 
\widetext
\begin{mathletters}
\label{Rdiv}
\begin{eqnarray}
P^{CD}_{AB}R_{CD}
&=&P^{CD}_{AB}\left[
-\frac{1}{2}D^aD_a\lambda_{CD}
+{1\over 4}(\lambda^{-1}D^a\lambda)D_a\lambda_{CD}
-\frac{1}{2}\lambda^{-1}c_C c_D
\right]
;
\label{e:R_AB_P}\\
\lambda^{AB}R_{AB}
&=&-\frac{1}{2}\lambda^{-1}D^aD_a\lambda
+{1\over 4}(\lambda^{-1}D^a\lambda)\lambda^{-1}D_a\lambda
-\frac{1}{2}\lambda^{-1}\lambda^{AB}c_A c_B
;
\label{e:R_AB_trace}\\
P^{cd}_{ab}\widehat{R}_{cd}
&=&
P^{cd}_{ab}\left[
-\frac{1}{2}\lambda^{-1}D_cD_d\lambda
+\frac{1}{4}\lambda^{-1}(D_c\lambda\lambda^{AB})(D_d\lambda_{AB})
+\frac{1}{4}\lambda^{-2}(D_c\lambda)(D_d\lambda)
\right]
;
\label{e:R_ab_P}
\\
\gamma^{ab}\widehat{R}_{ab}
&=&
-\frac{1}{2}\lambda^{-1}D^aD_a\lambda
+\frac{1}{4}\lambda^{-1}(D^a\lambda\lambda^{AB})(D_a\lambda_{AB})
+\frac{1}{4}\lambda^{-2}(D^a\lambda)(D_a\lambda)
\label{e:R_ab_trace}
+\mc{R}+\lambda^{-1}\lambda^{AB}c_A c_B
.
\end{eqnarray}
\end{mathletters}
\narrowtext

\subsection{Contracted Bianchi identities}
\label{ssec:bianchi}

Of course, all ten Einstein equations implied by Eq.~\eqref{e:R} are not
independent; they are related by the four contracted Bianchi
identities
\begin{equation}
  \label{prebianchi}
  \nabla^b G_{ab}=0
  .
\end{equation}
In this section, we express the various projections of
Eq.~\eqref{prebianchi} in terms of the projections
\begin{mathletters}
  \begin{eqnarray}
    G_{AB}&:=&K_A^c K_B^d G_{cd}
    \\
    \widehat{G}_{Ab}&:=&K_A^c\gamma_b^d G_{cd}
    \\
    \widehat{G}_{ab}&:=&\gamma_a^c\gamma_b^d G_{cd}
  \end{eqnarray}
\end{mathletters}
of the Einstein tensor.  Since $\mc{L}_{K_A}G_{ab}=0$, the symmetric
matrix $\{G_{AB}\}$ of scalar fields, the pair $\{\widehat{G}_{Ab}\}$
of covector fields, and the symmetric tensor field $\widehat{G}_{ab}$
all live on the two-manifold $\mc{S}$.  This means, in particular,
that
\begin{equation}
  \label{Gisperp}
  K_A^a\widehat{G}_{ab}=K_A^b\widehat{G}_{ab}=K_A^b\widehat{G}_{Ab}=0
\end{equation}
and, due to the vanishing of the Lie derivatives $\mc{L}_{K_C}G_{AB}$
and $\mc{L}_{K_B}\widehat{G}_{Ab}$,
\begin{mathletters}
  \label{liecons}
  \begin{eqnarray}
    \label{liescalar}
    K_C^c\nabla_c G_{AB}&=&0
    ;
    \\
    K_C^c\nabla_c \widehat{G}_{Ab}&=&-\widehat{G}_{Ac}\nabla_b K_C^c
    .
  \end{eqnarray}
\end{mathletters}

First, the projection of the contracted Bianchi identity along a
Killing vector is given by
\begin{eqnarray}
  0&=&K_A^a\nabla^b G_{ab}=\nabla^b (K_A^a G_{ab})
  \nonumber
  \\
  \nonumber
  &=&\nabla^b(\widehat{G}_{Ab}+\lambda^{BC}K_{Bb}G_{AC})
  =\nabla^b\widehat{G}_{Ab}
  \\
  \nonumber
  &=&D^b\widehat{G}_{Ab}+\widehat{G}_{Ac}\nabla^b\gamma_b^c
  =D^b\widehat{G}_{Ab}
  +\frac{1}{2}\widehat{G}_{Ab}\lambda^{-1}D^b\lambda
  ,
\\
\end{eqnarray}
where we used Eq.~\eqref{liescalar}, \eqref{e:div_gamma}, and the
antisymmetry of $\nabla^a K_B^b$ in $a$ and $b$.

Second, the projection \emph{orthogonal} to both Killing vectors is
\begin{eqnarray}
\nonumber
  0&=&\gamma_a^c\nabla^b G_{cb}
  =\gamma_a^d[\nabla^b(\gamma_d^c G_{cb})-G_{cb}\nabla^b\gamma_d^c]
  \\
  \nonumber
  &=&\gamma_a^d[\nabla^b(\widehat G_{db}
  +\lambda^{AB}K_{Bb}\widehat{G}_{Ad})
  +G_{cb}\lambda^{AB}K_A^c\nabla^b K_{Bd}]
  \\
  \nonumber
  &=&\gamma_a^d[\nabla^b\widehat G_{db}
  -\lambda^{AB}\widehat{G}_{Ab}\nabla_d K_B^b
  +\lambda^{AB}G_{Ab}\nabla^b K_{Bd}]
  \\
  \nonumber
  &=&\gamma_a^d[\nabla^b\widehat G_{db}
  -2\widehat{G}_{Ab}\lambda^{AB}\nabla_d K_B^b]
  +\frac{1}{2}G_{AB}D_a\lambda^{AB}
  .
  \\
  \label{g-bianchi}
\end{eqnarray}
The first term in Eq.~\eqref{g-bianchi} can be written as
\begin{eqnarray}
  \gamma_a^d\nabla^b\widehat{G}_{db}&=&D^b\widehat{G}_{ab}
   +\gamma_a^d \widehat{G}_{dc}\nabla^b\gamma_b^c
   \nonumber\\
  &=&D^b\widehat{G}_{ab}
  +\frac{1}{2}\widehat{G}_{ab}\lambda^{-1}D^b\lambda
  ,
\end{eqnarray}
where we again used Eq.~\eqref{g-bianchi}
\eqref{e:div_gamma} to simplify
$\nabla^b\gamma_b^c$.  Using Eq.~\eqref{e:grad_K} to replace the second
term gives the result
\begin{eqnarray}
  \nonumber
  0&=&\gamma_a^c\nabla^b G_{cb}
  =D^b\widehat{G}_{ab}
  +\frac{1}{2}\widehat{G}_{ab}\lambda^{-1}D^b\lambda
  +\frac{1}{2}G_{AB}D_a\lambda^{AB}
  \\
  &&+\epsilon^{-1}\abs{\lambda}^{-1/2}
  \epsilon_a{}^b\lambda^{AB}\widehat{G}_{Ab}c_B
  .
\end{eqnarray}
Defining the linear differential operator
\begin{equation}
  \mc{D}_a=D_a+\frac{1}{2}(\lambda^{-1}D_a\lambda)\times
  ,
\end{equation}
the contracted Bianchi identities can thus be written as
\begin{mathletters}
  \label{bianchi}
  \begin{eqnarray}
    \label{bianchi1}
    \mc{D}^b \widehat{G}_{Ab}&=&0
    ;
    \\
    \mc{D}^b\widehat{G}_{ab}&=&-\frac{1}{2}G_{AB}D_a\lambda^{AB}
    \nonumber\\
    \label{bianchi2}
    &&-\epsilon^{-1}\abs{\lambda}^{-1/2}
    \epsilon_a{}^b\lambda^{AB}\widehat{G}_{Ab}c_B
    .
  \end{eqnarray}
\end{mathletters}

The two relations \eqref{bianchi1} among the four off-block-diagonal
components $\{G_{Ai}\}$ of the Einstein tensor show why setting those
components to zero only allows us to eliminate two degrees of freedom
$\{c_A\}$ from the problem.

The two components of Eq.~\eqref{bianchi2} tell us that only four of the
remaining six components (the three $\{G_{AB}\}$ and the three
components $\{G_{ij}\}$ of $\widehat{G}_{ab}$) of the Einstein tensor
are independent.  Since Eqs.~\eqref{bianchi1}, \eqref{bianchi2}
are algebraic, rather than
differential, in the three projected components $\{G_{AB}\}$,
Eq.~\eqref{bianchi2} can be solved to give two of those components in
terms of the other eight.\footnote{This only works if the appropriate
  derivatives of $\lambda^{AB}$ are non-vanishing.  Otherwise, we have
  two differential relations among the components of the Einstein
  tensor.}

Also, Eq.~\eqref{bianchi2} can be rewritten, by substituting the form of
$\widehat{G}_{Ab}=\widehat{R}_{Ab}$ given by Eq.~\eqref{e:RKg}, as
\begin{equation}
  \mc{D}^b \widehat{G}_{ab}=-\frac{1}{2}G_{AB}D_a\lambda^{AB}
  +\frac{1}{2}\lambda^{-1}\lambda^{AB}c_A D_a c_B
  .
\end{equation}

\subsection{Recovering the four-geometry}
\label{ssec:4-geometry}

As shown in Sec.~\ref{ssec:prelim}, given a spacetime
$(\mc{M},g_{ab})$ admitting two commuting KVFs $\{K_A^a\}$, one can
define a number of tensor fields that live on the two-dimensional
space $\mc{S}$ of Killing vector orbits.  In particular, we defined
the two-metric $\gamma_{ab}$, the symmetric matrix of inner products
$\{\lambda_{AB}\}$, and the two scalar fields $\{c_A\}$.  In this
section, we complete our general discussion of spacetimes admitting
two KVFs by doing the converse.  That is, we show how to reconstruct
the four-geometry $(\mc{M},g_{ab})$ given only $\{\lambda_{AB}\}$,
$\{c_A\}$, and the metric components $\{\gamma_{ij}\}$ of
$\gamma_{ab}$ with respect to a {\coord} system $\{x^i\,|\,i=2,3\}$ on
$\mc{S}$.  The goal is to: (i) construct a basis
$\{e_\mu^a\,|\,\mu=0,1,2,3\}$ on $\mc{M}$; (ii) determine the
commutation {\coeff}s of this basis; (iii) specify the metric
components $\{g_{\mu\nu}\}$ of $g_{ab}$ with respect to this basis.
As we shall see below, if the scalar fields $c_A$ are non-zero, then
there is no preferred {\coord} basis on $\mc{M}$.
(Appendix~\ref{app:coord} describes the freedom in choosing a {\coord}
basis on $\mc{M}$ and the form the metric takes in such a basis.)
However, there is always a preferred \emph{non-{\coord}} basis on
$\mc{M}$, in terms of which the metric components $\{g_{\mu\nu}\}$ are
block-diagonal.

(i) Let $\{x^i\,|\, i=2,3\}$ be any {\coord} system on $\mc{S}$, and
let $\{\gamma_{ij}\}$ denote the components of $\gamma_{ab}$ with
respect to these {\coord}s---i.e.,
\begin{equation}
\gamma_{ab}=\gamma_{ij}(dx^i)_a(dx^j)_b\ .
\end{equation}
Then we can use the two contravariant vectors
\begin{mathletters}
\begin{equation}
e_i^a:=\gamma_{ij} g^{ab} (dx^j)_b\quad (i=2,3)
\label{e:e_i}
\end{equation}
along with the two Killing vector fields
\begin{equation}
e_A^a:=K_A^a\quad (A=0,1)
\label{e:e_A}
\end{equation}
\end{mathletters}
to define a basis $\{e_\mu^a\,|\,\mu=0,1,2,3\}$ on $\mc{M}$. 

Although the Killing vector fields commute with everything 
(i.e., $[e_A,e_\mu]^a=0$), and
\begin{equation}
[e_i,e_j]_{\mc{S}}^a:=e_i^b D_b e_j^a-e_j^b D_b e_i^a=
\gamma_b^a[e_i,e_j]^b=0\ ,
\end{equation}
the basis vectors $\{e_i^a\,|\,i=2,3\}$ need not commute on $\mc{M}$.
Thus, $\{e_\mu^a\,|\,\mu=0,1,2,3\}$ is not necessarily a {\coord}
basis on $\mc{M}$. 

(ii) Given a basis $\{e_\mu^a\,|\,\mu=0,1,2,3\}$ on $\mc{M}$, the 
commutation {\coeff}s $\{C_{\mu\nu}{}^\sigma\}$ are defined by
\begin{equation}
[e_\mu,e_\nu]^a=:C_{\mu\nu}{}^\sigma e_\sigma^a\ .
\end{equation}
Since, as mentioned in (i), 
\begin{equation}
[e_A,e_\mu]^a=0
\quad{\rm and}\quad
\gamma_b^a[e_i,e_j]^b=0\ ,
\end{equation}
it follows that
\begin{equation}
  [e_i,e_j]^a=C_{ij}{}^A e_A^a
\label{e:e_ie_j}
\end{equation}
define the only possible non-vanishing commutation {\coeff}s, which
are $\{C_{ij}{}^A\}$.

To calculate these {\coeff}s, invert Eq.~\eqref{e:e_ie_j}:
\begin{equation}
C_{ij}{}^A=[e_i,e_j]^a e_a^A=
[e_i,e_j]^a g_{ab}\lambda^{AB}K_B^b\ .
\end{equation}
Since $g_{ab}e_i^a K_B^b=0$ and $\nabla_a K_{Bb}=-\nabla_b K_{Ba}$, it
follows that
\begin{equation}
[e_i,e_j]^a g_{ab}K_B^b=-2e_i^a e_j^b\nabla_a K_{Ab}\ .
\end{equation}
Now use Eq.~\eqref{e:grad_K} to expand $\nabla_a K_{Ab}$.  The final
result is
\begin{equation}
C_{ij}{}^A=\epsilon^{-1}\abs{\lambda}^{-1/2}\epsilon_{ij}
\lambda^{AB}c_B\ ,
\label{e:C_23^A}
\end{equation}
where
\begin{equation}
\epsilon_{ij}:=e_i^a e_j^b\epsilon_{ab}
\end{equation}
are the components of the Levi-Civita tensor on $\mc{S}$.
Thus, we see that $c_A=0$ are the necessary and sufficient conditions
for $\{e_\mu^a|\mu=0,1,2,3\}$ to be a coordinate basis on $\mc{M}$.

(iii) Determining the metric components $g_{\mu\nu}$ with respect
to the basis $\{e_\mu^a\,|\,\mu=0,1,2,3\}$ is relatively 
straightforward.
Using the definitions (\ref{e:e_i}) and (\ref{e:e_A}), it follows that
\begin{mathletters}
\begin{eqnarray}
g_{AB}&:=& g_{ab}e_A^a e_B^b=\lambda_{AB}\ ,\\
g_{iB}&:=& g_{ab}e_i^a e_B^b=0\ ,\\
g_{ij}&:=& g_{ab}e_i^a e_j^b=\gamma_{ij}\ ,
\end{eqnarray}
\end{mathletters}
so the metric components
\begin{equation}
\{g_{\mu\nu}\}=
\left(
\begin{array}{cc}
\{\lambda_{AB}\} & 0\\
0 & \{\gamma_{ij}\}
\end{array}
\right)\ .
\end{equation}
are block diagonal.

\section{Further gauge fixing}
\label{sec:fix}

Within the formalism of Sec.~\ref{sec:2KVF}, there are still choices
to be made in defining a basis.  There is of course the choice of a
basis ({\coord} or otherwise) on the two-manifold $\mc{S}$, and we
will discuss possible {\coord} choices in Sec.~\ref{ssec:ijgauge}.
But it is also possible, by considering linear combinations of the
Killing vectors, to describe the same spacetime with different values
for $\{\lambda_{AB}\}$ and $\{c_A\}$, as we show in
Sec.~\ref{ssec:ABgauge}.

\subsection{Relabeling the Killing vectors}
\label{ssec:ABgauge}

The properties of the vectors $\{K_A^a\}$ which allow us to perform
the construction of Sec.~\ref{sec:2KVF} are that they obey Killing's
equation
\begin{mathletters}
  \begin{equation}
    \nabla^{(a}K_A^{b)}=0
    ,
  \end{equation}
  and that they commute with one another
  \begin{equation}
    K_A^b\nabla_{b}K_B^a-K_B^b\nabla_{b}K_A^a=0
    .
  \end{equation}
\end{mathletters}
If we define a new pair of vectors $\{K_{A'}^a\}$ to be a linear
combination of the first two:
\begin{equation}
  \label{Ktrans}
  K_{A'}^a=L_{A'}{}^B K_B^a
  ,
\end{equation}
then the new set of vectors will also be commuting Killing vectors if
and only if
\begin{mathletters}
  \label{primecond}
  \begin{equation}
    0=\nabla^{(a}K_{A'}^{b)}=K_B^{(a}\nabla^{b)}L_{A'}{}^B
  \end{equation}
  and
  \begin{eqnarray}
    0&=&
    K_{A'}^b\nabla_{b}K_{B'}^a-K_{B'}^b\nabla_{b}K_{A'}^a
    \nonumber
    \\
    &=&
    K_{A'}^b\nabla_{b}L_{B'}{}^C-K_{B'}^b\nabla_{b}L_{A'}{}^C
    .
  \end{eqnarray}
\end{mathletters}
Clearly a sufficient condition is that $\{L_{A'}{}^B\}$ be
constants.  It is also straightforward to show that if neither
$\lambda_{AB}$ nor $\lambda_{A'B'}$ is degenerate, it is also a
necessary condition.%
\footnote{This means that the linear combinations
must be constant except on the surfaces of signature change, and 
we are not interested in discontinuous transformations.}

Under this global $GL(2,\R)$ symmetry, $\lambda_{AB}$ transforms as a
second-rank covariant tensor, $\lambda$ as a scalar density of weight
two, and $c_A$ as a covariant vector density of weight one:
\begin{mathletters}
  \begin{eqnarray}
    \lambda_{A'B'}&=&L_{A'}{}^C L_{B'}{}^D \lambda_{CD} \\
    \lambda'&=&(\det L)^2 \lambda\\
    c'_{A'}&=&(\det L) L_{A'}{}^B c_B
    .
  \end{eqnarray}
\end{mathletters}
In particular, the value of
\begin{equation}
  I=\lambda^{-1}\lambda^{AB}c_Ac_B
\end{equation}
at a given point cannot be changed by the transformation
\eqref{Ktrans}, nor can the sign of $\lambda$.

We can use these transformations to bring $\lambda_{AB}$ and $c_A$
into a convenient form at one point in the two-manifold
$\mc{S}$---i.e., on one of the Killing vector orbits.  (In the case of
two equal-mass orbiting cosmic strings, where there is an additional
discrete rotational symmetry which exchanges the strings, a special
point is the fixed point of that rotation, which is the rotational
axis.)  The desired form depends on the invariant signs of $I$ and
$\lambda$:

(i) If $\lambda>0$, $\lambda_{AB}$ is positive definite, and thus $I$
must also be positive (or else the system of KVFs would be
orthogonally transitive).  We can choose the Killing vectors so that
$\lambda_{AB}=\delta_{AB}$ at our desired point, and use the residual
$SO(2)$ symmetry (which preserves that form of $\lambda_{AB}$) to
rotate $c_A$ so that $c_1=0$ and $c_0=\sqrt{I}$.

(ii) As mentioned in (i), $\lambda>0$ and $I<0$ is not allowed.

(iii) If $\lambda<0$ we can bring $\lambda_{AB}$ into a Lorentz form
$\lambda_{AB}=\eta_{AB}$.
If $I>0$, we define the KVFs so that $\lambda_{00}=-1$ and
$\lambda_{11}=1$ at our chosen point, and then use the residual
$SO(1,1)$ symmetry to set $c_1=0$ and $c_0=\sqrt{I}$.

(iv) If $\lambda<0$ and $I<0$, we instead define the KVFs so that 
$\lambda_{00}=1$ and
$\lambda_{11}=-1$ at the chosen point, and then use the residual
$SO(1,1)$ symmetry to set $c_1=0$ and $c_0=\sqrt{-I}$.

So, ignoring the special case where $I=0$, we always have the freedom
to set $c_1=0$ at a point.  Note that in the case of a vacuum
spacetime, where Eq.~\eqref{e:RKg} tells us that the $\{c_A\}$ are
constants, this means that $c_1$ vanishes everywhere. 

\subsection{Coordinate choices on the two-manifold}
\label{ssec:ijgauge}

The description in terms of the two-manifold $\mc{S}$ of Killing
vector orbits has been entirely {\coord}-independent, as emphasized by
the use of abstract index notation.  Thus we need to make a choice of
{\coord}s on $\mc{S}$ to complete the specification of a basis on the
four-manifold $\mc{M}$.  In vacuum spacetimes with orthogonal
transitivity ($c_A=0$), \eqref{e:R_AB_trace} implies that
\begin{equation}
  \label{divlambda}
  D^aD_a\sqrt{\pm\lambda}=0
  ,
\end{equation}
so (in the absence of signature change) $\sqrt{\abs{\lambda}}$ is a
harmonic {\coord} on $\mc{S}$ \cite{wald}.  Using
$\sqrt{\abs{\lambda}}$, along with its harmonic conjugate, leads to a
two-metric described only by a single conformal factor, and
effectively reduces the number of degrees of freedom in $\lambda_{AB}$
from three to two.  This is used, for example, to define the $\rho$
and $z$ (Weyl) {\coord}s in a general stationary,
axisymmetric, vacuum spacetime.

However, in the case at hand, $\lambda$ does not provide us with
a harmonic {\coord}, and thus this method does not work.  We can use
either a set of conformal {\coord}s \emph{or} a set based on
$\lambda$, as described in the following sections, but not both.

\subsubsection{Conformal {\coord}s}
\label{sssec:conf}

Since any two-manifold is conformally flat, we are of course free to
choose {\coord}s on $\mc{S}$ so that the metric is determined only by
a single conformal factor $\Phi(x^i)$:
\begin{equation}
  d\Sigma^2=\Phi\,
  \left[
    (dx^2)^2-\epsilon (dx^3)^2
  \right]
  .
\end{equation}
In this case, we would need two fields (constants \textit{in vacuo})
$c_0$ and $c_1$, to determine the commutation {\coeff}s for the
non-{\coord} basis described in Sec.~\ref{ssec:4-geometry}, plus three
more fields $\lambda_{00}$, $\lambda_{01}$ and $\lambda_{11}$ to
determine the $\{g_{AB}\}$ block of the metric, plus a single field
$\Phi$ to determine the $\{g_{ij}\}$ block of the metric, for a total
of six degrees of freedom [four \textit{in vacuo} if Eq.~\eqref{e:RKg} are
imposed \textit{a priori}], as summarized in Table~\ref{tab:1}.
\mediumtext
\begin{table}
\caption{Degrees of freedom in the absence of $\zeta$-symmetry.  
  This table enumerates the independent elements of $\{c_A\}$ and
  $\{\lambda_{AB}\}$ and the components $\{\gamma_{ij}\}$ of the
  two-metric for various {\coord} choices on the two-manifold $\mc{S}$
  of Killing vectors.  The fields $\{c_A\}$ become constants
  \textit{in vacuo}, or more generally when the off-block-diagonal
  components $\{T_{Ai}\}$ of the stress-energy tensor vanish, so both
  the $c_A$ and total counts are considered separately ``w/matter''
  (actually $T_{Ai}\ne0$) and ``\textit{in vacuo}'' (actually
  $T_{Ai}=0$).  The different columns correspond to the
  $\lambda$-based {\coord}s described in Sec.~\ref{sssec:lambda}, the
  conformal {\coord}s described in Sec.~\ref{sssec:conf}, and, for
  comparison, the {\coord}s which can be defined in vacuum
  ($T_{ab}=0$) spacetimes in the presence of orthogonal transitivity,
  which are \emph{both} $\lambda$-based \emph{and} conformal.  (The
  counting for the geodesic polar {\coord}s defined in
  Sec.~\ref{sssec:geod} is the same as for conformal {\coord}s, with
  the conformal factor $\Phi$ replaced by the metric component
  $\gamma_{33}$.)}
\label{tab:1}
  \begin{tabular}{c|ccc}
    &$\lambda$-based&Conformal&Weyl\tablenotemark[1] \\
    \tableline
    $c_A$ (w/matter) & 2 fields $c_0$,$c_1$ & 2 fields $c_0$,$c_1$ 
& N/A\tablenotemark[1] \\
    $c_A$ (\textit{in vacuo}) & 1 constant $c_0$ & 1 constant $c_0$
& 0 \\
    $\lambda_{AB}$ & 2 fields $\lambda_{01}$, $\lambda_{11}$ 
& 3 fields $\lambda_{00}$, $\lambda_{01}$, $\lambda_{11}$
& 2 fields $\lambda_{01}$, $\lambda_{11}$\\
    $\gamma_{ij}$ & 2 components $\gamma_{22}$, $\gamma_{33}$
& 1 conf.~fact.\ $\Phi$ & 1 conf.~fact.\ $\Phi$\\
    \tableline
    Total w/matter & 6 DOF & 6 DOF & N/A\tablenotemark[1] \\
    Total \textit{in vacuo} & 4 DOF + 1 param.\ 
& 4 DOF + 1 param.\ & 3 DOF
  \end{tabular}
  \tablenotetext[1]{Weyl coordinates, which are $\lambda$-based
    \emph{and} conformal, are only possible in vacuum spacetimes with
    orthogonal transitivity, and are included here for comparison
    only.}
\end{table}
\narrowtext

Note, however, that since the form of the flat metric depends on the
signature of the two-manifold $\mc{S}$ (Euclidean for $\epsilon<0$ and
Lorentzian for $\epsilon>0$), we would need to use different {\coord}
patches on either side of the signature change, and there would be no
meaningful relations between the definitions of $\Phi$ on either side.

\subsubsection{$\lambda$-based {\coord}s}
\label{sssec:lambda}

Although $\sqrt{\abs{\lambda}}$ is not a harmonic {\coord} in our
case, we can still reduce the number of independent components in the
metric by basing a {\coord} system on it.  We set $x^2=\lambda$ [in
the absence of Eq.~\eqref{divlambda}, there is no reason to work with the
square root], and choose $x^3$ so that the metric is diagonal, which
we can always do in two dimensions.

\begin{figure}
\begin{center}
\epsfig{file=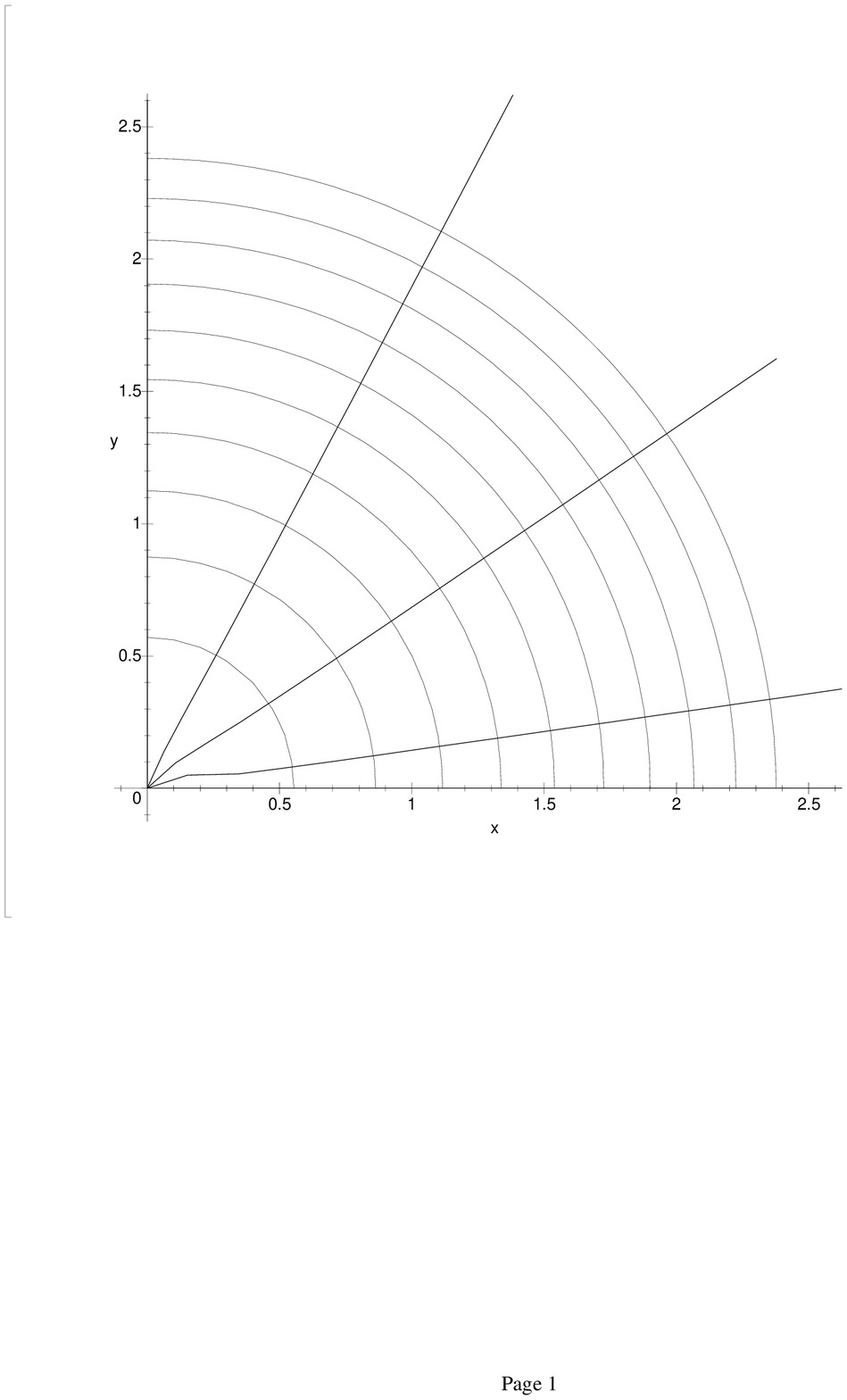,clip=,width=8.6cm,
bbllx=108,bblly=328,bburx=458,bbury=678}
\end{center}
  \caption{Surfaces of constant $\lambda$ on the two-manifold $\mc{S}$ and
    the surfaces orthogonal to those which are defined to have
    constant $\psi$.  At the center of rotation, $\lambda=-1$; the
    innermost level surface is for $\lambda=-0.9$, and the $\lambda$
    values increase by $0.1$ up to the outermost one, which is
    $\lambda=0$ (the light cylinder).  The constant-$\psi$ curves
    which are drawn are $\psi=0$ (the positive $x$-axis),
    $\psi=\pi/8$, $\psi=\pi/4$, $\psi=3\pi/8$, and $\psi=\pi/2$ (the
    positive $y$-axis).  This diagram was made by superposing two
    low-mass Levi-Civita cosmic strings (mass-per-unit-length $2C=0.2$
    in geometrical units) assumed to move in Newtonian orbit about one
    another.  (See Sec.~\ref{sssec:geod}.)  The {\coord}s drawn on the
    axes are the Cartesian analogues of polar {\coord}s $\rho$ and
    $\varphi$ in which the metric can be written in the form
    \eqref{geodmetric}.  The cosmic strings are located at $y=0$,
    $x=\pm 0.189$, and $\lambda$ becomes a bad coordinate about $0.02$
    from each of the strings.  It should thus be necessary, in a
    $\lambda$-based {\coord} system, to model the strings with
    boundary conditions placed at least that far from the strings.}
  \label{fig:lambda}
\end{figure}

We expect $\lambda$ to act as a radial {\coord}, since as a geometric
quantity it must be preserved by the discrete rotational symmetry
which exchanges the two strings.  In particular, this means that
constant-$\lambda$ surfaces must be closed.  [There is also the
analogy of co-rotating flat spacetime (Sec.~\ref{sec:problem}), where
$\lambda$ is related to the traditional radial {\coord} $\rho$ by
Eq.~\eqref{corotlambda}.]  Because of the choices made in
Sec.~\ref{ssec:ABgauge}, we know that $\lambda=-1$ at the origin (the
fixed point of the discrete rotational symmetry).  And the light
cylinder is, by definition, the surface on which $\lambda=0$.
Considering these concentric surfaces of constant $\lambda$
(Fig.~\ref{fig:lambda})
we can draw surfaces everywhere orthogonal to these, which are
surfaces of constant $x^3=\psi$.  If we call the one passing through
one cosmic string $\psi=0$ and the one passing through the other
string $\psi=\pi$, then $\psi$ is an angular {\coord} with period
$2\pi$.  This {\coord} can be defined so that the discrete symmetry
discussed above corresponds to a rotation by $\pi$ in $\psi$.  The
metric on $\mc{S}$ can be written in terms of two independent
components:
\begin{equation}
  d\Sigma^2=\gamma_{22}\,d\lambda^2
  +\gamma_{33}\,d\psi^2
  .
\end{equation}
For example, the two-metric \eqref{corotmet} for co-rotating flat
spacetime can be written in this form (the $\varphi$ used in
Sec.~\ref{sec:problem} is the same as the present $\psi$) with
\begin{mathletters}
  \label{corotmetcomp}
  \begin{eqnarray}
    \gamma_{22}&=&\frac{1}{4\Omega^2(1+\lambda)}\\
    \gamma_{33}&=&-\frac{1+\lambda}{\Omega^2\lambda}
    .
  \end{eqnarray}
\end{mathletters}

We have not yet completely defined the {\coord}s on the two-manifold;
the labeling of the constant $\psi$ surfaces is still to be specified.
Put another way, we can make a redefinition $\psi'=f(\psi)$ which
preserves all of the properties thus far discussed of the {\coord}
system, so long as $f(0)=0$ and $f(\psi+\pi)=f(\psi)+\pi$.  An obvious
way to finish that specification is to choose a particular value of
$\lambda$ and decree that equal intervals in $\psi$ sweep out equal
distance along that constant-$\lambda$ curve, or $\gamma_{33,\psi}=0$
for that value of $\lambda$.  Given the specifications we made at the
origin in Sec.~\ref{ssec:ABgauge}, the most convenient place to make
this definition is in the limit $\lambda\rightarrow -1$, assuming that
limit exists.

Counting the degrees of freedom needed to specify the metric, we again
have two fields (or constants) $c_0$ and $c_1$; now we only need (say)
$\lambda_{11}$ and $\lambda_{01}$ as functions of $\lambda$ and $\psi$
to specify the matrix $\{\lambda_{AB}\}$, since we know
\begin{equation}
  \lambda_{00}=\frac{\lambda+(\lambda_{01})^2}{\lambda_{11}}
  ;
\end{equation}
and finally we need the two diagonal components $\gamma_{22}$ and
$\gamma_{33}$ to specify the metric on $\mc{S}$.  Again, we have six
independent degrees of freedom, or four if the $c_A$ are taken to be
constant.  (See Table~\ref{tab:1}.)

\subsubsection{Geodesic polar {\coord}s}
\label{sssec:geod}

The use of $\lambda$ as a {\coord} has a number of potential hazards.
In addition to a {\coord} singularity at the origin, it may also fail
to be monotonic as one moves out from the origin to infinity.  For
example, consider the co-rotating spacetime obtained by superposing
two low-mass Levi-Civita cosmic strings\cite{cs}, each with
mass-per-unit length $2C$ in gravitational units, where $C$ is a small
dimensionless parameter.  This spacetime will \emph{not} solve
Einstein's equations (it includes only Coulomb effects and not
gravitomagnetic or radiative ones) but it can be written in our
formalism.  There exist {\coord}s on the two-manifold in which the
two-metric is
\begin{equation}
  \label{geodmetric}
  d\Sigma^2 = d\rho^2 - \frac{\rho^2}{\lambda}d\varphi^2,
\end{equation}
and in terms of those {\coord}s, the strings are located at $\rho=R$
and $\varphi=0,\pi$ and the matrix of Killing vector inner products is
given by\footnote{Note that if $C=0$, this just reduces to
  \eqref{lambdaAB}.}
\begin{equation}
  \label{geodlambda}
  \{\lambda_{AB}\}=
  \left(
    \begin{array}{cc}
      -[(\rho_{+}\rho_{-})^{2C}-\Omega^2\rho^2] & 0     \\
      0                   & (\rho_{+}\rho_{-})^{-2C}
    \end{array}
  \right)
  ,
\end{equation}
so that $\lambda=-1+(\rho_{+}\rho_{-})^{-2C}\Omega^2\rho^2$, where
$\rho_{\pm}$ are the Cartesian distances from each string, whose
product is
\begin{equation}
  (\rho_{+}\rho_{-})^2=(\rho^2+R^2)^2-4R^2\rho^2\cos^2\varphi
  .
\end{equation}
This means that sufficiently close to either of the strings, $\lambda$
becomes arbitrarily large, producing isolated (for small $C$) regions
around the strings which violate the assumption that $\lambda$
increases monotonically away from the origin.  We may be able to solve
this problem by removing those regions from the two-manifold and
modelling the strings by conditions on the resulting boundaries.

Note that the {\coord}s $\rho$ and $\varphi$ in which the metric has
been written have no such problems.  One way to generalize these
{\coord}s to arbitrary spacetimes would be to require
$\gamma_{22}\equiv 1$ and $\gamma_{23}\equiv 0$ everywhere.  This can
be achieved, for instance, by defining polar coordinates at the origin
and then requiring constant-$\varphi$ lines%
\footnote{Note that the angular {\coord} $\varphi$ in this system will
  not in general be the same as the $\psi$ defined in $\lambda$-based
  {\coord}s.}  to be geodesics along which $\rho$ is an affine
parameter, whence the name {\it geodesic polar {\coord}s}.  Then, just
as in the case of conformal {\coord}s, we will need only one function
($\gamma_{33}$ or equivalently $-\gamma_{33}\lambda/\rho^2$) to define
the two-metric, and three more to define the matrix $\lambda_{AB}$.

\subsection{An additional discrete symmetry}
\label{ssec:zeta}

This section does not actually concern gauge-fixing, but it does
describe a way in which the number of independent degrees of freedom
can be further reduced under certain circumstances.  Consider a
transformation which changes the sign of the Killing vector $K_1^a$
and simultaneously reverses the orientation of the spacetime $\mc{M}$.%
\footnote{Equation~\eqref{e:eps_ab} tells us that changing the sign of
  one of the Killing vectors must change the orientation of either
  $\mc{M}$ or $\mc{S}$.}  Under such a transformation, $c_1$,
$\lambda_{01}$, $R_{01}$, and $\{R_{1i}\}$ will change sign, but the
other parts of those objects, such as $c_0$, $\lambda_{00}$,
$\lambda_{11}$, etc., will not.  (In general, any object or component
will be transformed to $(-1)^\zeta$ times itself, where $\zeta$ is the
number of times $1$ appears as an index.)  Call this transformation
$\zeta$-reflection.

Now suppose we have a solution to the Einstein equations determined by
some stress-energy distribution and boundary conditions.  Note that
the auxiliary conditions on $\lambda_{AB}$ and $c_A$ at a single point
in $\mc{S}$, which we used to define the Killing vector labels, are
unchanged by $\zeta$-reflection (since they set $\lambda_{01}$ and
$c_1$ to zero at that point).  Thus $\zeta$-reflection of our initial
solution must also satisfy the Einstein equations, only with
$\zeta$-reflected stress-energy and boundary conditions.  If the
stress-energy and boundary conditions are sent into themselves by
$\zeta$-reflection (i.e., if their $\zeta$-odd components all vanish),
then we have two solutions to the same boundary-value problem, which
are $\zeta$-reflections of one another.  Assuming that the boundary
conditions used are sufficient to specify a unique solution, that
means that we are actually talking about one solution which is taken
into itself by $\zeta$-reflection, which means that the $\zeta$-odd
quantity $\lambda_{01}$ vanishes \emph{everywhere} in this
case.\footnote{The same can be concluded about $c_1$, but if the
  stress-energy is $\zeta$-even, it is already guaranteed to be a
  constant by \eqref{e:RKg}.}  (See Table~\ref{tab:2}.)
\mediumtext
\begin{table}
\caption{Degrees of freedom in the presence of $\zeta$-symmetry.
  This table enumerates the independent components of the quantities
  describing the four-geometry, just as in Table~\ref{tab:1}, but this
  time after we have assumed that the spacetime is unchanged by the
  $\zeta$-reflection defined in Sec.~\ref{ssec:zeta}, and so for
  example $\lambda_{01}\equiv 0$.  (The counting for the geodesic
  polar {\coord}s defined in Sec.~\ref{sssec:geod} is the same as for
  conformal {\coord}s, with the conformal factor $\Phi$ replaced by
  the metric component $\gamma_{33}$.)}
\label{tab:2}
  \begin{tabular}{c|ccc}
    &$\lambda$-based&Conformal&Weyl\tablenotemark[1]\\
    \tableline
    $c_A$ (w/matter) & 1 field $c_0$ & 1 field $c_0$
& N/A\tablenotemark[1] \\
    $c_A$ (\textit{in vacuo}) & 1 constant $c_0$ & 1 constant $c_0$
& 0 \\
    $\lambda_{AB}$ & 1 field $\lambda_{11}$ 
& 2 fields $\lambda_{00}$, $\lambda_{11}$
& 1 field $\lambda_{11}$\\
    $\gamma_{ij}$ & 2 components $\gamma_{22}$, $\gamma_{33}$
& 1 conf.~fact.\ $\Phi$ & 1 conf.~fact.\ $\Phi$\\
    \tableline
    Total w/matter & 4 DOF & 4 DOF & N/A\tablenotemark[1] \\
    Total \textit{in vacuo} & 3 DOF + 1 param.\ 
& 3 DOF + 1 param.\ & 2 DOF
  \end{tabular}
  \tablenotetext[1]{See footnote, table~\ref{tab:1}.}
\end{table}
\narrowtext

When are the stress-energy and boundary conditions going to be
$\zeta$-even?  The $\zeta$-odd parts $T_{01}$ and $T_{1i}$ of the
stress-energy can be though of as energy fluxes and shears along the
string, and setting them to zero would seem to be reasonable.
Considering the boundary conditions at infinity as describing
linearized radiation on a cylindrical background \cite{cyl}, and
thinking qualitatively in something like the transverse, traceless
gauge describing radiation moving radially outward, the ``plus''
polarization will involve the components $h_{\phi\phi}$ and $h_{zz}$
of the metric perturbation, which are $\zeta$-even quantities, while
the ``cross'' polarization will involve the $\zeta$-odd component
$h_{z\phi}$.  Thus $\zeta$-even boundary values are those which
involve only one polarization.

\section{The differential equations for co-rotating cosmic strings}
\label{sec:PDEs}

\subsection{General considerations}
\label{ssec:PDEgen}

\subsubsection{Number of degrees of freedom}
\label{sssec:DOF}

In a general spacetime, the Einstein equations are ten second-order,
non-linear, partial differential equations for the ten independent
metric components, as functions of the four spacetime {\coord}s.  As
described in Sec.~\ref{ssec:ijgauge}, after all of the gauge degrees
of freedom have been fixed out of a two-Killing-vector vacuum
spacetime (and $\widehat{G}_{Ab}$ have been set to zero \textit{a
  priori}), there remain four independent functions of the two
{\coord}s on the two-dimensional space of Killing vector orbits.
There are three Einstein equations involving $G_{AB}$ (conveniently
divided into the two involving the trace-free part
$P_{AB}^{CD}G_{CD}=P_{AB}^{CD}R_{CD}$ and one involving the trace
$\lambda^{AB}G_{AB}=-\gamma^{ab}\widehat{R}_{ab}$) and three more
involving $\widehat{G}_{ab}$ (two with
$P_{ab}^{cd}\widehat{G}_{cd}=P_{ab}^{cd}\widehat{R}_{cd}$ and one with
$\gamma^{ab}\widehat{G}_{ab}=-\lambda^{AB}R_{AB}$) for a total of six.%
\footnote{The contracted Bianchi identities, considered in
  Sec.~\ref{ssec:bianchi}, mean that only four of those six components
  are independent.  This is discussed further in
  Sec.~\ref{ssec:PDEspec}.}

If the spacetime is assumed to be unchanged by $\zeta$-reflection (see
Sec.~\ref{ssec:zeta}), there are only three independent degrees of
freedom in the gauge-fixed metric.  Since $G_{01}\equiv0$ in that
case, there are only five Einstein equations
involving those three functions of two variables.%
\footnote{The ``missing'' equation is one of the two involving
$P_{AB}^{CD}G_{CD}=P_{AB}^{CD}R_{CD}$.} 

\subsubsection{Order of the equations}

An important practical consideration for a numerical solution to the
Einstein equations is their order---i.e., the highest number of
derivatives appearing in each equation.  The Einstein equations for a
generic spacetime are second-order, but we will see that not all of
the equations enumerated in Sec.~\ref{sssec:DOF} are actually
second-order when $\lambda$ is used as one of the {\coord}s.  This is
because both of the components
\begin{equation}
  D_i\lambda\equiv\partial_i x^2=\delta_i^2
\end{equation}
are constant, and thus second covariant derivatives of $\lambda$ do
not translate into second derivatives of the functions of $\lambda$
and $\psi$ which define the metric:
\begin{equation}
  \label{DDl}
  D_iD_j\lambda=-\mc{G}_{ji}^2
  ,
\end{equation}
where
\begin{equation}
  \mc{G}_{ij}^k:=\frac{\gamma^{kl}}{2}
  (\gamma_{il,j}+\gamma_{jl,i}-\gamma_{ij,l})
\end{equation}
is a Christoffel symbol for the metric $\gamma_{ab}$ in the $\{x^i\}$
{\coord}s, and only involves first derivatives of metric components.

Thus, the only terms in Eq.~\eqref{Rdiv} which can involve second
derivatives of metric components are the scalar curvature (of
$\gamma_{ab}$) $\mc{R}$ and the trace-free (on $A$ and $B$) part of
$D_aD_b\lambda_{AB}$.  Defining the notation $\cong$ to mean
``equal up to first (and lower) derivative terms,'' we see that the
six Einstein equations divide into three second-order equations
involving
\begin{mathletters}
  \begin{eqnarray}
    \nonumber
    P_{AB}^{CD}G_{CD}&=&P_{AB}^{CD}R_{CD}
    \cong-\frac{1}{2}P_{AB}^{CD}D^aD_a\lambda_{CD}
    \\
    &\cong& -\frac{1}{2}D^aD_a\lambda_{AB}
    \\
    \lambda^{AB}G_{AB}&=&-\gamma^{ab}\widehat{R}_{ab}\cong-\mc{R}
  \end{eqnarray}
\end{mathletters}
and three
first-order equations involving 
\begin{mathletters}
  \begin{eqnarray}
    &&\gamma^{ab}\widehat{G}_{ab}=\lambda^{AB}R_{AB}\cong0
    \\
    &&P_{ab}^{cd}\widehat{G}_{cd}=P_{ab}^{cd}\widehat{R}_{cd}\cong0
    .
  \end{eqnarray}
\end{mathletters}

If the spacetime is assumed to be $\zeta$-symmetric,
$P_{01}^{CD}R_{CD}$ vanishes automatically, and there are only two
second-order equations and three first-order ones.

\subsection{Explicit forms}
\label{ssec:PDEspec}

\subsubsection{$\lambda$-based {\coord}s}

To determine the explicit differential equations obeyed by functions
defining the metric, it is convenient to use as the four independent
degrees of freedom
\begin{mathletters}
  \begin{eqnarray}
    X(\lambda,\psi)&:=&\lambda_{01}
    \\
    Z(\lambda,\psi)&:=&\lambda_{11}
    \\
    P(\lambda,\psi)&:=&\gamma_{22}
    \\
    h(\lambda,\psi)&:=&-\lambda\gamma_{33}/\gamma_{22}
    .
  \end{eqnarray}
\end{mathletters}
These definitions are chosen in part because they are all non-singular
at the light cylinder $\lambda=0$ in co-rotating flat spacetime
[cf.\ Eq.~~\eqref{corotmetcomp}]:
\begin{mathletters}
  \begin{eqnarray}
    X(\lambda,\psi)&=&0
    \\
    Z(\lambda,\psi)&=&1
    \\
    P(\lambda,\psi)&=&[4\Omega^2(1+\lambda)]^{-1}
    \\
    h(\lambda,\psi)&=&4(1+\lambda)^2
    .
  \end{eqnarray}
\end{mathletters}
Also $X$ is a $\zeta$-odd quantity, while the other three are
$\zeta$-even.%
\footnote{Note that while this does mean $X\equiv0$ in a
  $\zeta$-symmetric spacetime, the non-linearity of the Einstein
  equations will couple the $\zeta$-odd and $\zeta$-even parts of the
  metric.}

Substituting the expressions
\begin{mathletters}
  \begin{eqnarray}
    \lambda_{00}&=&(\lambda+X^2)Z^{-1}
    \\
    \lambda_{01}&=&X
    \\
    \lambda_{11}&=&Z
    \\
    \gamma_{22}&=&P
    \\
    \gamma_{23}&=&0
    \\
    \gamma_{33}&=&-\lambda^{-1}hP
    \\
    c_0&\equiv& 2\Omega
    \\
    c_1&\equiv& 0
  \end{eqnarray}
\end{mathletters}
into Eqs.~\eqref{e:R_AB_trace} and \eqref{e:R_ab_P}, a straightforward but
lengthy algebraic calculation 
gives the first-order equations
\widetext
\begin{mathletters}
  \label{Gij}
  \begin{eqnarray}
    -\gamma^{ab}\widehat{G}_{ab}=\lambda^{AB}R_{AB}
    &=&
   -\frac{1}{4}\lambda^{-1}h^{-1}P^{-1}h_\lambda
    -2\lambda^{-2}Z\Omega^2+\frac{1}{2}\lambda^{-2}P^{-1}
    ;
    \\
    \nonumber
    \widehat{G}_{23}=\widehat{R}_{23}
    \nonumber
    &=&-\frac{1}{2}(1+\lambda^{-1}X^2)Z^{-2}Z_\lambda Z_\psi
    -\frac{1}{2}\lambda^{-1}X_\lambda X_\psi
    +\frac{1}{2}\lambda^{-1}Z^{-1}
    X(Z_\lambda X_\psi+Z_\psi X_\lambda)
    \\
    &&
    +\frac{1}{4}\lambda^{-1}P^{-1}P_\psi
    +\frac{1}{4}\lambda^{-1}Z^{-1}Z_\psi
    ;
    \\
    \nonumber
    \gamma^{22}\widehat{G}_{22}-\gamma^{33}\widehat{G}_{33}
    &=&\gamma^{22}\widehat{R}_{22}-\gamma^{33}\widehat{R}_{33}
    \\
    \nonumber
    &=&-\frac{1}{2}(1+\lambda^{-1} X^{2})P^{-1}Z^{-2}Z_\lambda{}^2
    -\frac{1}{2}\lambda(1+\lambda^{-1} X^{2})
    h^{-1}P^{-1}Z^{-2}Z_\psi{}^2
    \\
    \nonumber
    &&+\lambda^{-1}P^{-1}Z^{-1}XZ_\lambda X_\lambda
    +h^{-1}P^{-1}Z^{-1}XZ_\psi X_\psi
    -\frac{1}{2}\lambda^{-1}P^{-1}X_\lambda{}^2
    \\
    &&-\frac{1}{2}h^{-1}P^{-1}X_\psi{}^2
    +\frac{1}{4}\lambda^{-1}h^{-1}P^{-1}h_\lambda
    +\frac{1}{2}\lambda^{-1}P^{-2}P_\lambda
    +\frac{1}{2}\lambda^{-1}P^{-1}Z^{-1}Z_\lambda
    ,
  \end{eqnarray}
\end{mathletters}
\narrowtext
where we have defined the shorthand
\begin{equation}
  h_\lambda:=\frac{\partial h}{\partial\lambda},\ 
  X_{\psi\psi}:=\frac{\partial^2 X}{\partial\psi^2},
  \text{ etc.}
\end{equation}

The contracted Bianchi identities derived in Sec.~\ref{ssec:bianchi}
mean that two of the second-order expressions $G_{AB}$ can be written
in terms of other components of $G_{ab}$.  Specifically, we can use
Eq.~\eqref{bianchi2} to solve for $G_{00}$ and $G_{11}$ as linear
combinations of $G_{01}$, $\{G_{ij}\}$, and $\{G_{ij,k}\}$.  The
explicit forms of those expressions are not needed to solve the
Einstein equations (since they simply show that $G_{00}$ and $G_{11}$
vanish identically in a vacuum).  However, the corresponding
expressions for $T_{00}$ and $T_{11}$ show that those components are
completely determined by the other components of the stress-energy
tensor $T_{ab}$ (and their derivatives) due to conservation of energy.

If the spacetime is assumed to be unchanged by $\zeta$-symmetry
(cf.~Sec.~\ref{ssec:zeta}), then $G_{01}\equiv 0$ automatically, and
the only independent Einstein equations are $G_{ij}=8\pi T_{ij}$, with
the components of $G_{ij}$ given by Eq.~\eqref{Gij}.  (Note that those
expressions are then further simplified by the condition $X\equiv 0$.)
In that case those three first-order equations, together with an
appropriate choice of boundary conditions, determine the three
functions $h(\lambda,\psi)$, $P(\lambda,\psi)$, and $Z(\lambda,\psi)$.

If $\zeta$-symmetry is not imposed, we have one other independent
component of the Einstein tensor, which we take to be
$P_{01}^{AB}G_{AB}=P_{01}^{AB}R_{AB}$.  Again, a bit of algebra
converts Eq.~\eqref{e:R_AB_P} into
\widetext
\begin{eqnarray}
  \nonumber
  P_{01}^{AB}G_{AB}&=&G_{01}-\frac{1}{2}X\lambda^{AB}G_{AB}
  \\
  \nonumber
  &=&-\frac{1}{2}P^{-1}X_{\lambda\lambda}
  +\frac{1}{2}\lambda h^{-1}P^{-1}X_{\psi\psi}
  +\frac{1}{2}\lambda^{-1}P^{-1}XX_{\lambda}{}^2
  -\frac{1}{2}h^{-1}P^{-1}XX_{\psi}{}^2
  \\
  \nonumber
  &&
  -\lambda^{-1}P^{-1}Z^{-1}X^2 Z_{\lambda}X_{\lambda}
  -\frac{1}{4}h^{-1}P^{-1} h_{\lambda}X_{\lambda}
  +h^{-1}P^{-1}Z^{-1}X^2 Z_{\psi}X_{\psi}
  -\frac{1}{4}\lambda h^{-2}P^{-1} h_{\psi}X_{\psi}
  \\
  \nonumber
  &&
  +\frac{1}{2}(1+\lambda^{-1}X^2)P^{-1}Z^{-2}XZ_{\lambda}{}^2
  -\frac{1}{2}\lambda(1+\lambda^{-1}X^2)
  h^{-1}P^{-1}Z^{-2}XZ_{\psi}{}^2
  +\frac{1}{2}\lambda^{-1}P^{-1}X_\lambda
  \\
  \label{G01}
  &&
  -\frac{1}{2}\lambda^{-1}P^{-1}Z^{-1}XZ_{\lambda}
  +\frac{1}{2}X\gamma^{ab}\widehat{G}_{ab}
  .
\end{eqnarray}
\narrowtext
If the spacetime is not assumed to be $\zeta$-symmetric, the four
functions $h(\lambda,\psi)$, $P(\lambda,\psi)$, $Z(\lambda,\psi)$, and
$X(\lambda,\psi)$ are determined by the three first-order partial
differential equations (PDEs)
corresponding to Eq.~\eqref{Gij} and the second-order PDE coming from
Eq.~\eqref{G01}, along with an appropriate set of boundary conditions.

\subsubsection{Geodesic polar {\coord}s}

The convenient degrees of freedom for defining the metric in geodesic
polar {\coord}s are
\begin{mathletters}
  \begin{eqnarray}
    X(\rho,\varphi)&:=&\lambda_{01}
    \\
    Z(\rho,\varphi)&:=&\lambda_{11}
    \\
    \lambda(\rho,\varphi)&:=&
    \lambda_{00}\lambda_{11}-\lambda_{01}{}^2
    \\
    F(\rho,\varphi)&:=&-\lambda\gamma_{33}/\rho^2
    .
  \end{eqnarray}
\end{mathletters}
Now the light cylinder is a surface in $\rho,\varphi$ space defined by
$\lambda(\rho,\varphi)=0$, and the flat-space forms of the other three
quantities are well-behaved on that surface
[cf.\ Eqs.~(\ref{geodmetric}),(\ref{geodlambda}), with $C=0$]:
\begin{mathletters}
  \begin{eqnarray}
    X(\rho,\varphi)&=&0
    \\
    Z(\rho,\varphi)&=&1
    \\
    \lambda(\rho,\varphi)&=&-(1-\Omega^2\rho^2)
    \\
    F(\rho,\varphi)&=&1
    .
  \end{eqnarray}
\end{mathletters}
Note that the quantities $X$ and $Z$ are the same as those defined in
$\lambda$-based {\coord}s.

Substituting the expressions
\begin{mathletters}
  \begin{eqnarray}
    \lambda_{00}&=&(\lambda+X^2)Z^{-1}
    \\
    \lambda_{01}&=&X
    \\
    \lambda_{11}&=&Z
    \\
    \gamma_{22}&=&1
    \\
    \gamma_{23}&=&0
    \\
    \gamma_{33}&=&-\lambda^{-1}\rho^2 F
    \\
    c_0&\equiv& 2\Omega
    \\
    c_1&\equiv& 0
  \end{eqnarray}
\end{mathletters}
into Eqs.~\eqref{e:R_AB_trace} and \eqref{e:R_ab_P}, another
straightforward algebraic calculation gives the equations
\widetext
\begin{mathletters}
  \label{Gijgeod}
  \begin{eqnarray}
    -\gamma^{ab}\widehat{G}_{ab}=\lambda^{AB}R_{AB}
    &=&
    -\frac{1}{2}\lambda^{-1}\lambda_{\rho\rho}
    +\frac{1}{2}\rho^{-2}F^{-1}\lambda_{\varphi\varphi}
    +\frac{1}{2}\lambda^{-2}\lambda_\rho{}^2
    -\frac{1}{4}\lambda^{-1}F^{-1}\lambda_\rho F_\rho
    \\
    \nonumber
    &&
    -\frac{1}{4}\rho^{-2}F^{-2}\lambda_\varphi F_\varphi
    -\frac{1}{2}\rho^{-1}\lambda^{-1}\lambda_\rho
    -2\lambda^{-2}Z\Omega^2
    ;
    \\
    \widehat{G}_{23}=\widehat{R}_{23}
    &=&
    -\frac{1}{2}\lambda^{-1}\lambda_{\rho\varphi}    
    -\frac{1}{2}(1+\lambda^{-1}X^2)Z^{-2}Z_\rho Z_\varphi
    -\frac{1}{2}\lambda^{-1}X_\rho X_\varphi
    +\frac{1}{4}\lambda^{-1}F^{-1}F_\rho\lambda_\varphi
    \\
    \nonumber
    &&
    +\frac{1}{4}\lambda^{-1}Z^{-1}
    (\lambda_\rho Z_\varphi+\lambda_\varphi Z_\rho)
    +\frac{1}{2}\lambda^{-1}Z^{-1}X(Z_\rho X_\varphi+Z_\varphi X_\rho)
    +\frac{1}{2}\rho^{-1}\lambda^{-1}\lambda_\varphi
    ;
    \\
    \nonumber
    \gamma^{22}\widehat{G}_{22}-\gamma^{33}\widehat{G}_{33}
    &=&\gamma^{22}\widehat{R}_{22}-\gamma^{33}\widehat{R}_{33}
    \\
    &=&-\frac{1}{2}\lambda^{-1}\lambda_{\rho\rho}
    -\frac{1}{2}\rho^{-2}F^{-1}\lambda_{\varphi\varphi}
    +\frac{1}{4}\lambda^{-1}F^{-1}\lambda_\rho F_\rho
    +\frac{1}{4}\rho^{-2}F^{-2}\lambda_\varphi F_\varphi
    \\
    \nonumber
    &&
    +\frac{1}{2}\lambda^{-1}Z^{-1}\lambda_\rho Z_\rho
    +\frac{1}{2}\rho^{-2}F^{-1}Z^{-1}\lambda_\varphi Z_\varphi
    -\frac{1}{2}(1+\lambda^{-1} X^{2})Z^{-2}Z_\rho{}^2
    \\
    \nonumber
    &&
    -\frac{1}{2}\lambda(1+\lambda^{-1} X^{2})
    \rho^{-2}F^{-1}Z^{-2}Z_\varphi{}^2
    +\lambda^{-1}Z^{-1}XZ_\rho X_\rho
    +\rho^{-2}F^{-1}Z^{-1}XZ_\varphi X_\varphi
    \\
    \nonumber
    &&
    -\frac{1}{2}\lambda^{-1}X_\rho{}^2
    -\frac{1}{2}\rho^{-2}F^{-1}X_\varphi{}^2
    +\frac{1}{2}\rho^{-1}\lambda^{-1}\lambda_\rho
    ,
  \end{eqnarray}
\end{mathletters}
which again uses the shorthand
\begin{equation}
  h_\rho:=\frac{\partial h}{\partial\rho},\ 
  X_{\varphi\varphi}:=\frac{\partial^2 X}{\partial\varphi^2},
  \text{ etc.}
\end{equation}

Note that these equations are now second order in
$\lambda(\rho,\varphi)$ but not in the other dependent variables.  As
in the previous section, these three equations, along with appropriate
boundary conditions, are enough to specify the three functions
$\lambda$, $F$, and $Z$ if the spacetime is assumed to be even under
$\zeta$-symmetry (in which case $X$ vanishes everywhere).  If not, we
need a fourth, $\zeta$-odd, equation derived from Eq.~\eqref{e:R_AB_P}:
\begin{eqnarray}
  \label{G01geod}
  P_{01}^{AB}G_{AB}&=&G_{01}-\frac{1}{2}X\lambda^{AB}G_{AB}
  \\
  \nonumber
  &=&
  -\frac{1}{2}X_{\rho\rho}
  +\frac{1}{2}\lambda \rho^{-2}F^{-1}X_{\varphi\varphi}
  +\frac{1}{2}\lambda^{-1}XX_\rho{}^2
  -\frac{1}{2}\rho^{-2}F^{-1}XX_\varphi{}^2
  -\lambda^{-1}Z^{-1}X^2 Z_\rho X_\rho
  \\
  \nonumber
  &&
  +\rho^{-2}F^{-1}Z^{-1}X^2 Z_\varphi X_\varphi
  +\frac{1}{2}\lambda^{-1} \lambda_\rho X_\rho
  -\frac{1}{4}F^{-1} F_\rho X_\rho
  -\frac{1}{4}\rho^{-2}\lambda F^{-2} F_\varphi X_\varphi
  \\
  \nonumber
  &&
  +\frac{1}{2}(1+\lambda^{-1}X^2)Z^{-2}XZ_\rho{}^2
  -\frac{1}{2}\lambda(1+\lambda^{-1}X^2)
\rho^{-2}F^{-1}Z^{-2}XZ_\varphi{}^2
  -\frac{1}{2}\lambda^{-1}Z^{-1}X\lambda_\rho Z_\rho
  \\
  \nonumber
  &&
  +\frac{1}{2}\rho^{-2}F^{-1}Z^{-1}X\lambda_\varphi Z_\varphi
  -\frac{1}{2}\rho^{-1}X_\rho
  +\frac{1}{2}X\gamma^{ab}\widehat{G}_{ab}
  .
\end{eqnarray}
\narrowtext
For a spacetime without $\zeta$-symmetry, the four functions
$\lambda(\rho,\varphi)$, $F(\rho,\varphi)$, $Z(\rho,\varphi)$, and
$X(\rho,\varphi)$ are determined by the four second-order PDEs
corresponding to Eqs.~\eqref{Gijgeod} and \eqref{G01geod}, along with an
appropriate set of boundary conditions.

\section{Conclusions}
\label{sec:concl}

To summarize, we have used the two Killing vectors present in the
spacetime of a pair of co-rotating cosmic strings to help us find the
simplest set of quantities (four functions of two variables; three if
the spacetime is assumed to be $\zeta$-symmetric as defined in
Sec.~\ref{ssec:zeta}) needed to describe the geometry of that
spacetime.  Since we learned in Sec.~\ref{sec:problem} that it is not
possible (due to a lack of orthogonal transitivity) to define a
subspace everywhere orthogonal to both Killing vectors, we instead
worked on the two-dimensional space $\mc{S}$ of Killing vector orbits.

In Sec.~\ref{sec:2KVF}, we derived the components of the Einstein
tensor in terms of tensor fields on the two-manifold $\mc{S}$, thereby
streamlining and extending the derivation by Geroch \cite{GEROCH2} of
the vacuum Einstein equations.  We also expressed the contracted
Bianchi identities in terms of those components, and showed that two
components of the Einstein tensor (and not just their derivatives)
were (generically) simply linear combinations of the other components
and their derivatives.

In Sec.~\ref{sec:fix}, we described further gauge fixing possible
within the Geroch formalism; in particular, the choice of {\coord}s on
the two-manifold $\mc{S}$.  One possibility was to choose {\coord}s so
that the two-geometry was described only by a single conformal factor;
however, that was seen to be inconvenient due to the fact that
$\mc{S}$ had a Lorentzian signature outside of the ``light cylinder''.
A more convenient set of {\coord}s was found, which used the
determinant $\lambda$ of the inner products of the Killing vectors as
a radial {\coord}.  This reduced the number of independent scalar
fields needed to describe the four-geometry.  A third possibility is
to mimic polar {\coord}s by defining lines of constant angular
{\coord} to be spacelike geodesics in $\mc{S}$ radiating out from the
origin and using the distance along those curves as a radial {\coord}.
With any of these {\coord} choices, the spacetime geometry was found
to be described by one parameter (representing the frequency of the
fixed rotation) and four independent functions of the two {\coord}s on
$\mc{S}$, as detailed in Table~\ref{tab:1}.  If, as described in
Sec.~\ref{ssec:zeta}, a further discrete symmetry (called
$\zeta$-symmetry) was imposed upon the spacetime, the number of
independent degrees of freedom was reduced to three, as detailed in
Table~\ref{tab:2}.

Finally, in Sec.~\ref{sec:PDEs} we found explicit forms, in the
$\lambda$-based {\coord} system and the geodesic polar {\coord}
system, for the Einstein equations in terms of the four independent
functions described in the previous section.  This meant finding
expressions
for four components of the Einstein tensor.%
\footnote{Four of the ten components of the Einstein tensor $G_{ab}$
  had been set to zero \textit{a~priori}, and two more could be
  determined by the Bianchi identities derived in
  Sec.~\ref{ssec:bianchi}.}

It was shown that, in $\lambda$-based {\coord}s, three of those
components led to first-order partial differential equations---i.e.,
they contained no second derivatives of the functions which described
the metric.  The fourth independent component, which was second order,
vanished identically if the $\zeta$-symmetry of Sec.~\ref{ssec:zeta}
was imposed.  The Einstein equations were then written as a set of
four (homogeneous \textit{in vacuo}, inhomogeneous in the presence of
matter) partial differential equations, three of them first-order and
one second-order, for four functions of two variables.  If the sources
and boundary conditions uniquely specify a solution, and neither of
them breaks the discrete $\zeta$-symmetry, the system of PDEs further
simplifies to become three first-order equations for three functions
of two variables.

In geodesic polar coordinates, which quasi-Newtonian descriptions of
orbiting cosmic strings indicate may be better behaved, the situation
was analogous, except that second derivatives of one of the functions
appeared in all of the Einstein equations.

To conclude, we note two aspects of the problem that were outside 
the scope of this paper, but which will be addressed in future papers 
in this series:

(i) First, we said very little about the sources appearing in the
Einstein equations.  The spacetime is taken to be vacuum away from the
cosmic strings, but the stress-energy of the strings themselves (and
in fact whether they are better described by a distributional
stress-energy or a set of boundary conditions on a small circle
surrounding each string) is not discussed in this paper.  However,
since we always kept track of which components appeared in expressions
which were to be set to zero to give the vacuum Einstein equations, we
will be able to insert any sources once we have a stress-energy
tensor.

(ii) Second, we left aside the issue of boundary conditions, both at
infinity and at the axis of rotation.  In order to keep the system in
equilibrium, we expect to need a balance of incoming and outgoing
radiation in the exterior boundary condition.  How to implement such a
condition in our problem is a subject for further research
\cite{standing,cyl}.  In addition, there will be complications in
imposing boundary conditions at the axis of rotation (which
corresponds to $\lambda=-1$ in our {\coord}s), since the $\lambda$,
$\psi$ {\coord} system is badly singular there, as the metric
components \eqref{corotmetcomp} for co-rotating flat spacetime show.
This, along with the fact that $\lambda$ is not single-valued near the
strings in the quasi-Newtonian spacetime discussed in
Sec.~\ref{sssec:geod}, implies that the geodesic polar {\coord} system
will prove more convenient.

\section*{Acknowledgments}

The authors would like to thank R.~H.~Price and C.~Torre for many
useful discussions.  J.T.W.\ would also like to thank W.~Krivan,
K.~Thorne, J.~Creighton, J.~Friedmann, S.~Morsink, A.~Held, and the
Relativity Group at the University of Utah, where this work was begun.
J.T.W.\ was supported by NSF Grant PHY-9734871, by Swiss Nationalfonds,
and by the Tomalla Foundation Z\"{u}rich.  J.D.R.\ was supported by NSF
Grants PHY-9308728 and PHY-9503084.

\appendix

\section{Calculation of the covariant derivative\\
of the Killing vector fields}
\label{app:grad_K}

To calculate the covariant derivative of the Killing vector fields, we
start by noting that Killing's equation (\ref{e:KVF}) implies that
$\nabla_a K_{Ab}$ is antisymmetric in its $a$ and $b$ indices.  Thus,
$\nabla_a K_{Ab}$ is completely specified by its contractions with the
tensor fields $\epsilon^{ab}$, $\gamma_c^{[a}K_D^{b]}$, and
$K_C^{[a}K_D^{b]}$.

(i) The contraction of $\nabla_a K_{Ab}$ with $\epsilon^{ab}$ is 
straightforward.
Using Eqs.~(\ref{e:c_A}) and (\ref{e:eps_ab}), it follows that
\begin{eqnarray}
\nonumber
\epsilon^{ab}\nabla_a K_{Ab}&=&
\abs{\lambda}^{-1/2}\epsilon^{abcd}K_{0c}K_{1d}\nabla_a K_{Ab}
\\
&=&
\abs{\lambda}^{-1/2} c_A\ .
\label{e:contract1}
\end{eqnarray}

(ii) To obtain the contraction of $\nabla_a K_{Ab}$ with 
$\gamma_c^{[a}K_D^{b]}$, it is convenient to note that%
\footnote{This result follows by differentiating Eq.~\eqref{e:lambda_AB}.}
\begin{equation}
K_D^b\nabla_a K_{Ab}=\frac{1}{2}D_a\lambda_{AD}\ .
\label{e:D_lambda}
\end{equation}
Thus,
\begin{equation}
\gamma_c^{[a}K_D^{b]}\nabla_a K_{Ab}=\frac{1}{2}D_c\lambda_{AD}\ .
\label{e:contract2}
\end{equation}

(iii) Similarly, by using Eq.~(\ref{e:D_lambda}), we find
\begin{equation}
K_C^{[a}K_D^{b]}\nabla_a K_{Ab}=0\ .
\label{e:contract3}
\end{equation}

Since
\begin{equation}
-\frac{1}{2}\epsilon^{-1}\abs{\lambda}^{-1/2}
\epsilon_{ab}\ c_A
-\lambda^{BC}K_{B[a}D_{b]}\lambda_{CA}
\end{equation}
has the same three contractions with $\epsilon^{ab}$,
$\gamma_c^{[a}K_D^{b]}$, and $K_C^{[a}K_D^{b]}$ as does $\nabla_a
K_{Ab}$, we have proven the equality (\ref{e:grad_K}).

\section{Calculation of the projected\\
components of the  Ricci tensor}
\label{app:ricci}

The following three subsections contain proofs of 
Eqs.~(\ref{e:RKK}), (\ref{e:RKg}), and (\ref{e:Rgg}).
These results were stated without proof in the main text.

\subsection{Proof of Eq.~(\ref{e:RKK})}
\label{ssec:proof1}

Using the definition $R_{bd}:=R^c{}_{bcd}$ and Eqs.~(\ref{e:Riem_K}),
(\ref{e:D_lambda}), and (\ref{e:KVF}), it follows that
\begin{eqnarray}
\nonumber
R_{AB}&:=&K_A^aK_B^bR_{ab}=R^c{}_{acb}K_A^aK_B^b
=(\nabla_c\nabla_b K_A^c)K_B^b
\\
\nonumber
&=&\nabla_c(K_B^b\nabla_bK_A^c)-(\nabla_cK_B^b)(\nabla_b K_A^c)
\\
&=&-\frac{1}{2}\nabla_cD^c\lambda_{AB}+(\nabla_bK_{Ac})(\nabla^bK_B^c)
\ .
\label{e:3c1}
\end{eqnarray}
The first term on the right-hand side (RHS) of Eq.~(\ref{e:3c1}) can be
evaluated by writing $D^c\lambda_{AB}$ as $\gamma_d^cD^d\lambda_{AB}$,
and then differentiating $\gamma_d^c$ and $D^d\lambda_{AB}$
separately.  The result is
\begin{equation}
-\frac{1}{2}\nabla_c D^c\lambda_{AB}
=-{1\over 4}(\lambda^{-1}D_d\lambda)D^d\lambda_{AB}
-\frac{1}{2}D_d D^d\lambda_{AB}\ ,
\end{equation}
where we used
\begin{eqnarray}
\nonumber
\nabla_c\gamma_d^c
&=&\nabla_c(\delta_d^c-\lambda^{AB}K_A^cK_{Bd})
=-\lambda^{AB}K_A^c \nabla_cK_{Bd}
\\
&=&\frac{1}{2}\lambda^{AB}D_d\lambda_{AB}
=\frac{1}{2}\lambda^{-1}D_d\lambda\ .
\label{e:div_gamma}
\end{eqnarray}
The second term on the RHS of Eq.~(\ref{e:3c1}) can be evaluated by 
using Eq.~(\ref{e:grad_K}) to expand the covariant derivatives of the
Killing vector fields.
The two cross-terms, which are proportional to 
$\epsilon_{bc}K^{[b}D^{c]}$, vanish.
Only the $\epsilon_{bc}\epsilon^{bc}$ and $K_{[b}D_{c]}K^{[b}D^{c]}$ 
terms remain.
Explicitly,
\begin{eqnarray}
(\nabla_b K_{Ac})&&(\nabla^b K_B^c)
={1\over 4}\epsilon^{-2}\abs{\lambda}^{-1}
\epsilon_{bc}\epsilon^{bc}\ c_A c_B
\\
\nonumber
&&+\lambda^{CD}\lambda^{EF}K_{C[b}(D_{c]}\lambda_{DA})
K_E^{[b}(D^{c]}\lambda_{FB})
\\
  \nonumber
&&=-\frac{1}{2}\lambda^{-1}c_A c_B
+\frac{1}{2}\lambda^{DF}(D_c\lambda_{DA})(D^c\lambda_{FB})\ .
\end{eqnarray}
Thus
\begin{eqnarray}
\nonumber
R_{AB}&=&
-{1\over 4}(\lambda^{-1}D^a\lambda)D_a\lambda_{AB}
-\frac{1}{2}D^aD_a\lambda_{AB}
\\
&&-\frac{1}{2}\lambda^{-1}c_A c_B
+\frac{1}{2}\lambda^{CD}(D^a\lambda_{AC})(D_a\lambda_{BD})\ .
\label{e:RKK_1}
\end{eqnarray}
If desired, the RHS of this last expression can be rewritten using the 
identity
\begin{eqnarray}
  \label{e:matrix_id}
  \lambda^{CD}(D^a\lambda_{AC})(D_a\lambda_{BD})&=&
  (\lambda^{-1}D^a\lambda)D_a\lambda_{AB}
  \\
  \nonumber
  &&-\frac{1}{2}\lambda_{AB}\lambda^{-1}
  (D^a\lambda \lambda^{CD})D_a\lambda_{CD}
  ,
\end{eqnarray}
which holds for any invertible $2\times 2$ matrix $\lambda_{AB}$.
Substituting Eq.~(\ref{e:matrix_id}) into Eq.~(\ref{e:RKK_1}) yields
Eq.~\eqref{e:RKK}.

\subsection{Proof of Eq.~(\ref{e:RKg})}
\label{ssec:proof2}

To obtain the projected components 
$\widehat{R}_{Ab}:=K_A^c\gamma_b^d R_{cd}$, we start by writing
\begin{eqnarray}
\nonumber
\widehat{R}_{Ac}&=&K_A^a\gamma_c^bR_{ab}=R^d{}_{adb}K_A^a\gamma_c^b
=(\nabla_d\nabla_bK_A^d)\gamma_c^b
\\
&=&\gamma_c^a\nabla^b\nabla_a K_{Ab}
\ .
\end{eqnarray}
We then use Eq.~(\ref{e:grad_K}) to expand $\nabla_a K_{Ab}$:
\begin{eqnarray}
\nonumber
\gamma_c^a\nabla^b\nabla_a K_{Ab}
&=&\gamma_c^a\nabla^b\left(
-\frac{1}{2}\epsilon^{-1}\abs{\lambda}^{-1/2}\epsilon_{ab}\ c_A
\right.
\\
\nonumber
&&
\phantom{
  \gamma_c^a\nabla^b\left(\vphantom{\frac{1}{2}}\right.
  }
\left.
\vphantom{\frac{1}{2}}
-\lambda^{BC}K_{B[a}D_{b]}\lambda_{CA}\right)
\\
\nonumber
&=&-\frac{1}{2}\gamma_c^a\nabla^b
(\epsilon^{-1}\abs{\lambda}^{-1/2}\epsilon_{ab}\ c_A)
\\
\label{e:3c2}
&&
-\gamma_c^a\nabla^b(\lambda^{BC}K_{B[a}D_{b]}\lambda_{CA})\ .
\end{eqnarray}
A straightforward calculation shows that the 2nd term on the RHS of 
Eq.~(\ref{e:3c2}) vanishes, while the 1st term can be calculated by first 
writing $\epsilon^{-1}\abs{\lambda}^{-1/2}\epsilon_{ab}\ c_A$ as
$\epsilon^{-1}\abs{\lambda}^{-1/2}\gamma_b^d\ \epsilon_{ad}\ c_A$,
and then differentiating $\gamma_b^d$ and 
$\epsilon^{-1}\abs{\lambda}^{-1/2}\epsilon_{ad}\ c_A$ separately.
The result of this differentiation is that the first term equals
\begin{eqnarray}
\nonumber
{\rm 1st\  term}&=&
-{1\over 4}\epsilon^{-1}\abs{\lambda}^{-1/2}(\lambda^{-1}D^d\lambda)
\epsilon_{cd}\ c_A
\\
&&
-\frac{1}{2}D^d(\epsilon^{-1}\abs{\lambda}^{-1/2}\epsilon_{cd}\ c_A)
\ ,
\label{e:3c3}
\end{eqnarray}
where we used Eq.~(\ref{e:div_gamma}) and definition (\ref{e:D_a}).
Moreover, by differentiating each of the factors of
$\epsilon^{-1}\abs{\lambda}^{-1/2}\epsilon_{cd}\ c_A$ separately, we
find
\begin{eqnarray}
\label{e:3c4}
&&-\frac{1}{2}D^d(\epsilon^{-1}\abs{\lambda}^{-1/2}\epsilon_{cd}\ c_A)
\\
\nonumber
&&={1\over 4}\epsilon^{-1}\abs{\lambda}^{-1/2}(\lambda^{-1}D^d\lambda)
\epsilon_{cd}\ c_A
-\frac{1}{2}\epsilon^{-1}\abs{\lambda}^{-1/2}\epsilon_{cd}D^dc_A
,
\end{eqnarray}
where we used%
\footnote{If we allow $\epsilon$ to change discontinuously from $-1$
  to $+1$, $D_a\epsilon$ and $D_a\epsilon_{bc}$ do {\em not\,} vanish.
  They are related by Eq.~(\ref{e:D_epsilon}).}
\begin{equation}
D^d\epsilon_{cd}=\frac{1}{2}\epsilon_{cd}\ \epsilon^{-1}D^d\epsilon
\label{e:D_epsilon}
\end{equation}
to eliminate the derivatives of $\epsilon$ and $\epsilon_{cd}$.
Finally, by combining Eqs.~(\ref{e:3c3}) and (\ref{e:3c4}), and recalling 
that the 2nd term on the RHS of Eq.~(\ref{e:3c2}) vanishes, we get
Eq.~\eqref{e:RKg}.

\subsection{Proof of Eq.~(\ref{e:Rgg})}
\label{ssec:proof3}

To obtain the final set of projected components 
$\widehat{R}_{ab}:=\gamma^c_a\gamma^d_b R_{cd}$,
we proceed in a manner similar to Geroch \cite{GEROCH2}, and consider
the Riemann tensor $\mc{R}^a{}_{bcd}$ of the two-dimensional metric 
$\gamma_{ab}$:
\begin{equation}
\mc{R}^a{}_{bcd}v^b:=2D_{[c} D_{d]}v^a\ .
\label{e:def_cR}
\end{equation}
Here $D_a$ is the covariant derivative operator compatible with
$\gamma_{ab}$ [see Eq.~(\ref{e:D_a})], and $v^a$ is an arbitrary
vector field on $\mc{S}$---i.e., it satisfies
\begin{equation}
K_{Aa}v^a=0
\quad{\rm and}\quad
\mc{L}_{K_A}v^a=0\ .
\end{equation}
In particular, $K_{Aa}v^a=0$ implies $\nabla_a(K_{Ab}v^b)=0$, 
so that
\begin{equation}
K_A^b\nabla_a v_b=-v^b\nabla_a K_{Ab}\ ,
\label{e:3d1}
\end{equation}
while $\mc{L}_{K_A} v^a=0$ is equivalent to
\begin{equation}
K_A^a\nabla_a v_b=v^a\nabla_a K_{Ab}\ .
\label{e:3d2}
\end{equation}
Equations (\ref{e:3d1}) and (\ref{e:3d2}) will be needed below to relate 
$\mc{R}^a{}_{bcd}$ to the four-dimensional Riemann tensor
\begin{equation}
R^a{}_{bcd}v^b:=2\nabla_{[c} \nabla_{d]}v^a\ .
\end{equation}

Using the symmetry properties of the Riemann tensor, it follows that
Eq.~(\ref{e:def_cR}) is equivalent to
\begin{equation}
\mc{R}_{abcd}v^d=2D_{[a} D_{b]}v_c\ .
\end{equation}
Using Eq.~(\ref{e:D_a}) to evaluate the RHS, we find
\widetext
\begin{eqnarray}
\nonumber
2D_{[a} D_{b]}v_c
&=&2\gamma_{[a}^d\gamma_{b]}^e\gamma_c^f\nabla_d
(\gamma_e^g\gamma_f^h\nabla_g v_h)
\\
&=&2\gamma_{[a}^d\gamma_{b]}^e\gamma_c^f
\left[
(\nabla_d\gamma_e^g)\gamma_f^h\nabla_g v_h
+\gamma_e^g(\nabla_d\gamma_f^h)\nabla_g v_h
+\gamma_e^g\gamma_f^h\nabla_d\nabla_g v_h
\right]\ .
\end{eqnarray}
The last term above can be rewritten as
\begin{equation}
2\gamma_{[a}^d\gamma_{b]}^g\gamma_c^h\nabla_d\nabla_g v_h
=\gamma_{[a}^d\gamma_{b]}^g\gamma_c^h R_{dghf}v^f
=\gamma_{[a}^e\gamma_{b]}^f\gamma_c^g\gamma_d^h R_{efgh}v^d\ ,
\end{equation}
while the first two terms can be rewritten as
\begin{eqnarray}
\nonumber
\text{1st two terms}
&=&2\gamma_{[a}^d\gamma_{b]}^e\gamma_c^h
(\nabla_d\gamma_e^g)\nabla_g v_h
+2\gamma_{[a}^d\gamma_{b]}^g\gamma_c^f
(\nabla_d\gamma_f^h)\nabla_g v_h
\\
\nonumber
&=&-2\lambda^{AB}
\left[
\gamma_{[a}^d\gamma_{b]}^e\gamma_c^h
(\nabla_d K_{Ae})K_B^g\nabla_g v_h
+\gamma_{[a}^d\gamma_{b]}^g\gamma_c^f
(\nabla_d K_{Af})K_B^h\nabla_g v_h
\right]
\\
&=&-2\lambda^{AB}
\left[
\gamma_{[a}^d\gamma_{b]}^e\gamma_c^h
(\nabla_d K_{Ae})(\nabla_g K_{Bh})v^g
-\gamma_{[a}^d\gamma_{b]}^g\gamma_c^f
(\nabla_d K_{Af})(\nabla_g K_{Bh})v^h
\right]\ ,
\end{eqnarray}
where we used Eqs.~(\ref{e:3d1}) and (\ref{e:3d2}) to obtain the 
last equality.
After a little more ``index gymnastics,'' we find
\begin{equation}
  \text{1st two terms}=
2\lambda^{AB}
\gamma_{[a}^e\gamma_{b]}^f\gamma_c^g\gamma_d^h
\left[
(\nabla_e K_{Af})(\nabla_g K_{Bh})
+(\nabla_e K_{Ag})(\nabla_f K_{Bh})
\right]
v^d\ .
\end{equation}
Thus, since $v^d$ is arbitrary,
\begin{equation}
\mc{R}_{abcd}=
\gamma_{[a}^e\gamma_{b]}^f\gamma_c^g\gamma_d^h
\left[
R_{efgh}
+2\lambda^{AB}(\nabla_e K_{Af})(\nabla_g K_{Bh})
+2\lambda^{AB}(\nabla_e K_{Ag})(\nabla_f K_{Bh})
\right]\ .
\label{e:3d3}
\end{equation}

Now contract Eq.~(\ref{e:3d3}) with $\gamma^{ac}$.  The left-hand side
is simply $\mc{R}_{bd}$, while the 1st term on the RHS is given by
\begin{eqnarray}
\nonumber
\gamma_b^f\gamma_d^h\gamma^{eg}R_{efgh}
&=&\gamma_b^f\gamma_d^h\left(\delta^{eg}
  -\lambda^{AB}K_A^e K_B^g\right)
R_{efgh}
\\
\nonumber
&=&\gamma_b^a\gamma_d^c
\left(R_{ac}-\lambda^{AB}K_A^e\nabla_a\nabla_e K_{Bc}\right)
\\
\nonumber
&=&\gamma_b^a\gamma_d^c
\left[R_{ac}-\lambda^{AB}\nabla_a(K_A^e\nabla_e K_{Bc})
+\lambda^{AB}(\nabla_a K_A^e)\nabla_e K_{Bc}\right]
\\
\label{e:3d4}
&=&\gamma_b^a\gamma_d^c
\left[R_{ac}+\frac{1}{2}\lambda^{AB}(D_a D_c\lambda_{AB})
+\lambda^{AB}(\nabla_a K_A^e)\nabla_e K_{Bc}\right]\ .
\end{eqnarray}
\narrowtext
By substituting Eq.~(\ref{e:3d4}) into Eq.~(\ref{e:3d3}),
and using Eq.~(\ref{e:grad_K}) to expand the covariant derivatives
of the Killing vector fields, we find
\begin{eqnarray}
\nonumber
\mc{R}_{bd}&=&
\gamma^a_b \gamma^c_d R_{ac}
+\frac{1}{4}\lambda^{AB}D_bD_d\lambda_{AB}
+\frac{1}{4}D_b(\lambda^{-1}D_d\lambda)
\\
&&
-\frac{1}{2}\gamma_{bd}\lambda^{-1}\lambda^{AB}c_A c_B
\ ,
\end{eqnarray}
which is equivalent to
\begin{eqnarray}
\label{e:temp}
\widehat{R}_{ab}:=\gamma^c_a \gamma^d_b R_{cd}&=&
\mc{R}_{ab}
-\frac{1}{4}\lambda^{AB}D_aD_b\lambda_{AB}
\\
\nonumber
&&
-\frac{1}{4}D_a(\lambda^{-1}D_b\lambda)
+\frac{1}{2}\gamma_{ab}\lambda^{-1}\lambda^{AB}c_A c_B
.
\end{eqnarray}
If desired, the RHS of this last expression can be simplified further 
using the identity
\begin{equation}
\mc{R}_{abcd}=\mc{R}\gamma_{a[c}\gamma_{d]b}
,
\end{equation}
which holds in two dimensions.  
In particular,
\begin{equation}
\mc{R}_{ab}=\frac{1}{2}\mc{R}\gamma_{ab}
\end{equation}
(i.e., the two-dimensional Einstein tensor vanishes), so that
\begin{eqnarray}
\nonumber
\widehat{R}_{ab}&=&
-\frac{1}{4}\lambda^{AB}D_aD_b\lambda_{AB}
-\frac{1}{4}D_a(\lambda^{-1}D_b\lambda)
\\
\label{Rgg1}
&&
+\frac{1}{2}\gamma_{ab}
(\mc{R}+\lambda^{-1}\lambda^{AB}c_A c_B)
.
\end{eqnarray}
In addition, it will prove useful to use the identity
\begin{eqnarray}
  \nonumber
  \lambda^{AB}D_aD_b\lambda_{AB}&=&\lambda^{-1}D_aD_b\lambda
  \\
  \label{lDDl}
  &&
  -\lambda^{-1}(D_a\lambda\lambda^{AB})(D_b\lambda_{AB})
\end{eqnarray}
to convert Eq.~\eqref{Rgg1} into the final form, Eq.~\eqref{e:Rgg}

\section{Coordinate bases on the four-manifold}
\label{app:coord}

In Sec.~\ref{ssec:4-geometry} we demonstrated that the metric tensor
components and commutation {\coeff}s in a non-{\coord} basis on the
four-manifold $\mc{M}$ can be obtained from a set of fields on the
two-manifold $\mc{S}$ of Killing vector orbits: the scalar fields
$\{\lambda_{AB}\}$ and $\{c_A\}$ and the components $\{\gamma_{ij}\}$ in a
coordinate system $\{x^i\}$ of the metric tensor $\gamma_{ab}$.  For those
uncomfortable with this definition of ``recovering the four-geometry,'' we
discuss in this appendix the definition of a \emph{\coord} basis on $\mc{M}$,
and the components of the metric in that basis.

In assigning {\coord}s $\{x^\mu|\mu=0,1,2,3\}$ to every point in the
spacetime $\mc{M}$, we have two things at our disposal:

The {\coord}s $\{x^i|i=2,3\}$ on the manifold $\mc{S}$ of Killing
vector orbits, which can be used to assign values of $\{x^i\}$ to each
Killing trajectory, and hence to each point along each Killing
trajectory.  This definition means that the basis one-forms
$e_a^i=(dx^i)_a$ on $\mc{S}$ are also two of the four basis
one-forms on $\mc{M}$.

The Killing vectors $\{K^a_A|A=0,1\}$, which can be used to assign
values of $\{x^A\}$ relative to some origin, along each Killing vector
orbit (which is itself a two-dimensional subspace of $\mc{M}$).  This
definition means that the directional derivative along the Killing
vector $K^a_A$ is $\partial/\partial x^A$, and gives us two basis
vectors $e_A^a=K_A^a$ on $\mc{M}$.

However, the assignment of all four coordinates to each point on
$\mc{M}$ is not complete until we state where on each Killing
trajectory the origin $x^0=x^1=0$ is, and in particular how that
origin is carried from one Killing trajectory to the next.
Identifying these surfaces of constant $x^A$ is equivalent to defining
the two basis one-forms\footnote{We use the tilde to distinguish this
  {\coord} basis from the non-{\coord} basis in
  Sec.~\ref{ssec:4-geometry}.} $\widetilde{e}_a^A=(dx^A)_a$.  We know
their components along the Killing vectors are $\widetilde{e}_a^A
K_B^a=\partial x^A/\partial x^B=\delta^A_B$, but their components
orthogonal to the Killing vectors give us two as yet unknown one-forms
$\{\beta_a^A\}$ on $\mc{S}$:
\begin{equation}
  \label{betadef}
  \beta_a^A:= (dx^A)_a-\lambda^{AB}K_{Ba}
  .
\end{equation}

The specification of the $\{\beta_a^A\}$ completes the definition of a
{\coord} system on $\mc{M}$ (up to an overall additive constant in
$x^A$), and we can explicitly write the implied definitions of the
complementary basis one-forms $\{\widetilde{e}_a^A\}$ and vectors
$\{\widetilde{e}_i^a\}$, as well as the covariant and contravariant
components of the metric tensor.  To do this, we first note that since
the directional derivative along $\widetilde{e}_i^a$ is
$\partial/\partial x^i$, we know that $\widetilde{e}_i^a
e^j_a=\delta_i^j$.  This means that any vector \emph{on $\mc{S}$} will
have the same components along $e_i^a$ as it does along
$e_i^a=\gamma_{ij}g^{ab}(dx^j)_b$, the latter form being defined
independent of the specification of $\{\beta_a^A\}$.  The components
$\beta_i^A=e_i^a \beta_a^A$ are thus defined in a
non-circular way, and it is reasonable to write the basis forms and
vectors and metric tensor components in terms of $\{\beta_i^A\}$:
\begin{mathletters}
  \begin{eqnarray}
    \widetilde{e}_a^A&=&\lambda^{AB}K_{Ba}+\beta_i^A e_a^i
    \\
    \label{eiA}
    \widetilde{e}_i^a&=&e_i^a-\beta_i^A K_A^a
    \\
    \widetilde{g}_{AB}&=&\lambda_{AB}
    \\
    \label{gAj}
    \widetilde{g}_{Aj}&=&-\lambda_{AB}\beta^B_j
    \\
    \label{gij}
    \widetilde{g}_{ij}&=&\gamma_{ij}+\lambda_{AB}\beta^A_i\beta^B_j
    \\
    \widetilde{g}^{AB}&=&\lambda^{AB}+\gamma^{ij}\beta^A_i\beta^B_j
    \\
    \widetilde{g}^{iB}&=&\gamma^{ij}\beta^B_j
    \\
    \widetilde{g}^{ij}&=&\gamma^{ij}
    .
  \end{eqnarray}
\end{mathletters}

We know that $\{\beta^A_a\}$ depend on the choice of
constant-$\{x^A\}$ surfaces, but we have not so far specified what
values are allowed for the functions $\{\beta^A_i\}$, nor have we made
use of the parameters $\{c_A\}$ which we know prevent us from setting
$\{\beta^A_a\}$ to zero.  We do both now by noting that we found in
Sec.~\ref{ssec:4-geometry} that
\begin{equation}
[e_i,e_j]^a=\epsilon^{-1}\abs{\lambda}^{-1/2}\epsilon_{ij}
\lambda^{AB}c_B K_A^a
\end{equation}
[cf.\ Eq.~\eqref{e:C_23^A}].  On the other hand,
Eq.~\eqref{eiA}, along with the
condition that $\{\widetilde{e}_i^a\}$ be part of a {\coord} basis on
$\mc{M}$, tells us that
\begin{equation}
  [e_i,e_j]^a=(\beta^A_{j,i}-\beta^A_{i,j})K_A^a
  .
\end{equation}
Combining the two gives
\begin{equation}
  \label{curlbeta}
  \epsilon^{ab}\partial_a\beta^A_b=-\abs{\lambda}^{-1/2}\lambda^{AB}c_B
  ,
\end{equation}
which is the equation which must be obeyed by $\{\beta_a^A\}$.  Of
course, this condition does not completely specify the one-forms
$\{\beta_a^A\}$; the left-hand side is unchanged by the ``gauge
transformation''
\begin{equation}
  \beta^A_a\rightarrow \beta^A_a + \partial_a \xi^A
\end{equation}
where $\{\xi^A\}$ are arbitrary scalar fields on $\mc{S}$.  Not
coincidentally, we see from the definition \eqref{betadef} that this
is \emph{exactly} the change induced in $\{\beta_a^A\}$ by a
displacement (not constant over $\mc{S}$)
\begin{equation}
  x^A\rightarrow x^A +  \xi^A
\end{equation}
in the {\coord}s along the Killing trajectories.

One possible gauge fixing would be to define geodesic polar {\coord}s
(Sec.~\ref{sssec:geod})
on $\mc{S}$ and then impose the condition $\beta^A_\rho=0$, which
allows $\beta^A_\varphi$ to be defined by integrating \eqref{curlbeta}
with respect to $\rho$.  We see from Eq.~\eqref{gAj} and Eq.~\eqref{gij} that
$\widetilde{g}_{A\rho}=0$ and $\widetilde{g}_{i\rho}=\delta_i^\rho$, 
which is just the normal
form (29.1) specified in \cite{petrov} for the metric in a {\coord}
basis on a four-manifold with a two-dimensional Abelian symmetry
group.

So, in order to completely specify the components of the metric tensor
in a {\coord} basis, we must be provided, in addition to the
quantities listed in Sec.~\ref{ssec:4-geometry}, with four additional
functions $\{\beta_i^A\}$ to replace the two functions $\{c_A\}$.
However, the information in those functions beyond that given by
$\{c_A\}$ is simply gauge information needed to state which definition
of a constant-$x^A$ surface we are using.

\section{Correspondence with the formulas of \protect\cite{GEROCH2}}
\label{app:geroch}

While the derivation, in Sec.~\ref{sec:2KVF}, of the components of the
Einstein tensor $G_{ab}$ does not exactly parallel that used by Geroch
in \cite{GEROCH2} to find the vacuum Einstein equations, the purpose
of this appendix is to point out the counterparts of the equations in
Geroch's Appendix~A, where they exist, and to present the equivalent
equations in our notation when they do not.  One thing to note is that
in many cases two or three equations in Geroch can be written as a
single equation in our notation, since we use the indices
$A$, $B$, etc.\ to discuss the behavior of quantities like the matrix
$\{\lambda_{AB}\}$ of inner products, while Geroch lists the
components $\lambda_{00}$, $\lambda_{01}$ and $\lambda_{11}$
separately.

Geroch's conditions (A1),(A2) for a tensor to be on $\mc{S}$ are our
Eqs.~(\ref{e:orthog}),(\ref{e:Lie_drag}).  Note that we write the Killing
vectors as $\{K_A^a|A=0,1\}$, while he writes them as
$\stackrel{0}{\xi^a}$ and $\stackrel{1}{\xi^a}$.  Geroch's definition
(A3) of $\lambda_{AB}$ is our Eq.~\eqref{e:lambda_AB}, and his (A4) is our
Eq.~\eqref{e:det}.  Note that the determinant $\lambda$ is written in his
notation as $-\tau^2/2$ (since he assumes it to be negative).  His
definition (A5) is our Eq.~\eqref{e:gamma_ab}, with our $\gamma_{ab}$
being the same as his $h_{ab}$.  His (A6) is our Eq.~\eqref{e:eps_ab}, or,
written without resort to the use of explicit values for the $A$ and
$B$ indices, Eq.~\eqref{e:eps_ab_new}.  His definition (A7) of the
covariant derivative on $\mc{S}$ becomes our Eq.~\eqref{e:D_a}.  As noted
in Sec.~\ref{ssec:proj_ricci}, we do not define a matrix of twist
vectors in our derivation, but we can write Geroch's definitions (A8)
of the twists compactly as
\begin{equation}
  \omega_{AB}^a=\epsilon^{abcd}K_{(Ab}\nabla_c K_{B)d}.
\end{equation}
[We have here defined the convention that symmetrization (and
antisymmetrization) operate independently on abstract indices like
$a,b,\ldots$ and concrete indices like $A,B,\ldots$ .]  Geroch's
definitions (A9) of $c_0$ and $c_1$ are our Eq.~\eqref{e:c_A} or--more
elegantly written--Eq.~\eqref{e:c_A_new}.  The projections (A10) of the
twists are of course written
\begin{equation}
  \nu_{AB}^a=\gamma^a_b\omega_{AB}^b
  .
\end{equation}
Geroch makes the statements, between displayed equations (A10) and
(A11), that the vacuum Einstein equations indicate that $c_0$ and
$c_1$ are constants and that the twists are curl-free.  In the
non-vacuum case, the former statement becomes our Eq.~\eqref{e:RKg}, and
the latter becomes
\begin{equation}
  \nabla_{[a}\omega_{ABb]}=-\epsilon_{abcd}R^{c}_e K_{(A}^{d}K_{B)}^e
  .
\end{equation}
The curls (A11) of the projected twists are written in our notation as
\begin{equation}
  D_{[a}\nu_{ABb]}=\epsilon_{ab}\epsilon_{(A}{}^CK_{B)}^d R_{cd}K_C^c
  +\frac{\epsilon_{(A}{}^C c_{B)}c_C}{2\lambda}\epsilon_{ab}
  ,
\end{equation}
where we have used $\lambda^{AB}$ to raise the second index on
$\epsilon_{AB}$.  Geroch's (A12) is one component of our
Eq.~\eqref{e:D_lambda}; his (A13) is the one pair of equations which
cannot be cannot be written in a compact form because they only
involve the norms of the Killing vectors ($\lambda_{00}$ and
$\lambda_{11}$) and not their inner product ($\lambda_{01}$), and
similarly only the diagonal elements of the twist matrix
$\omega_{AB}^a$.  This is because each is basically a result from the
formalism with only one Killing vector\cite{geroch1}. There is,
however, an analogous equation
\begin{eqnarray}
  \nonumber
  \nabla_a K_{Cb}=\frac{3}{4}
  \left(\vphantom{\frac{1}{2}}\right.
  &&
  \frac{1}{2}\lambda^{AB}\epsilon_{abcd}K_{(A}^c\omega_{BC)}^d
  \\
  \label{curlKomega}
  &&\left.\vphantom{\frac{1}{2}}
    -\lambda^{AB}K_{(C[a}\nabla_{b]}\lambda_{AB)}
  \right)
\end{eqnarray}
for the covariant derivative of a Killing vector in terms of the
twists.  [Of course, by deriving Eq.~\eqref{e:grad_K} initially, our
derivation has circumvented these steps.]  Geroch's equations (A14)
can also be derived from Eq.~\eqref{curlKomega}, and written as
\begin{equation}
  \nu_{AB}^a=-\epsilon^{ab}\epsilon_{(A}{}^C D_b\lambda_{B)C}
  .
\end{equation}
Geroch's first Einstein equations (A15) are
equivalent to the trace-free equation \eqref{e:R_AB_P}; specifically,
they arise from setting to zero in turn the components of
$\epsilon_{(A}{}^C R_{B)C}$.  The fact that only two of these three
equations are independent is apparent from the fact that they involve
only the trace-free part of the matrix $\{R_{AB}\}$.  Geroch obtains a
third independent equation (A16) by calculating $D^aD_a\lambda_{00}$;
one can equivalently look at $\lambda^{AB}D^aD_a\lambda_{AB}$, which
leads to the additional equation \eqref{e:R_AB_trace}.  The three
independent equations, which Geroch writes as (A18), are just the
components of Eq.~\eqref{e:RKK}.  Geroch obtains a fourth Einstein
equation by starting from an equation (A20) for the Riemann tensor on
$\mc{S}$ which can be written in our compact notation as
\begin{equation}
  \label{Riemtwo}
  \mc{R}_{abcd}=\gamma_{[a}^e\gamma_{b]}^f\gamma_c^g\gamma_d^h
  \left[
    R_{efgh}+4\lambda^{AB}(\nabla_e K_{A(f})(\nabla_{g)} K_{Bh})
  \right]
  .
\end{equation}
Now, Geroch contracts with $\gamma^{ab}$ on both pairs of indices to
obtain the equation (A21), which is equivalent to our
Eq.~\eqref{e:R_ab_trace}.  In so doing, however, he misses the equation
\eqref{e:R_ab_P} for the trace-free part of $\widehat{R}_{ab}$.  For,
while it is true that all three terms in Eq.~\eqref{Riemtwo} have the
symmetries of the Riemann tensor and are thus completely characterized
by their trace on the two-dimensional manifold $\mc{S}$, when the
first term on the right-hand side is traced over one set of indices to
give $\gamma_a^c\gamma_b^d\gamma^{ef}R_{cedf}$, this manifestly
trace-only quantity is split into $\gamma_a^c\gamma_b^d R_{cd}$ and
another piece, each of which may individually have non-vanishing
trace-free parts.

Note that while Geroch only provided four of the six block-diagonal
vacuum Einstein equations, the contracted Bianchi identities described
in Sec.~\ref{ssec:bianchi} mean that only four of these six equations
are independent anyway.  However, the identities are algebraic only in
the components $G_{AB}$, and the ``missing'' equations are for
$P_{ab}^{cd}\widehat{R}_{cd}=P_{ab}^{cd}\widehat{G}_{cd}$.  The
statement about those components implied by the Bianchi identity
\eqref{bianchi2} (setting the off-block-diagonal components of the
Einstein tensor to zero identically) is
\begin{equation}
  \mc{D}^b P_{ab}^{cd}\widehat{R}_{ab}
  =\mc{D}^b (\gamma_{ab}\lambda^{AB}R_{AB})
  -\frac{1}{2}G_{AB}D_a\lambda^{AB}
  ,
\end{equation}
which seems to imply that two of Geroch's four equations are
higher-order than they need to be.  (That is, they are implied by the
derivatives of equations he leaves out.)

\end{document}